\documentclass[%
reprint,
amsmath,amssymb,  aps, prc, nofootinbib
]{revtex4-1}
\usepackage{mathtools}
\usepackage{graphicx}
\usepackage{dcolumn}
\usepackage{bm}
\usepackage{subfigure} 
\usepackage{xcolor}

\begin{document}

\title{Impact of hadronic interactions and conservation laws on cumulants of conserved charges in a dynamical model}

\author{Jan~Hammelmann$^{1,2}$ and Hannah~Elfner$^{3,2,1,4}$}

\affiliation{$^1$Frankfurt Institute for Advanced Studies, Ruth-Moufang-Strasse 1, 60438
Frankfurt am Main, Germany}
\affiliation{$^2$Institute for Theoretical Physics, Goethe University,
Max-von-Laue-Strasse 1, 60438 Frankfurt am Main, Germany}
\affiliation{$^3$GSI Helmholtzzentrum f\"ur Schwerionenforschung, Planckstr. 1, 64291
Darmstadt, Germany}
\affiliation{$^4$Helmholtz Research Academy Hesse for FAIR (HFHF), GSI Helmholtz Center,
Campus Frankfurt, Max-von-Laue-Straße 12, 60438 Frankfurt am Main, Germany}

\keywords{Hadron gas, fluctuations}
\date{\today}

\begin{abstract}
Understanding the phase diagram of QCD by measuring fluctuations of conserved charges in heavy-ion collision is one of the main goals of the beam energy scan program at RHIC.
Within this work, we calculate the role of hadronic interactions and momentum cuts on cumulants of conserved charges up to fourth order in a system in equilibrium within a hadronic transport approach (SMASH).
In our model the net baryon, net charge and net strangeness is perfectly conserved on an event-by-event basis and the cumulants are calculated as a function of subvolume sizes and compared to analytic expectations.
We find a modification of the kurtosis due to charge annihilation processes in systems with simplified degrees of freedom. Furthermore the result of the full SMASH hadron gas for the net baryon and net proton number fluctuations
is presented for systems with zero and finite values of baryon chemical potential. Additionally we find that due to dynamical correlations
the cumulants of the net baryon number cannot easily be recovered from the net protons. Finally the influence of deuteron cluster formation on the net proton and net baryon fluctuations in a simplified system is shown.
This analysis is important to better understand the relation between measurements of fluctuations in heavy-ion collisions and theoretical calculation which are often performed in a grand canonical ensemble.
\end{abstract}

\maketitle

\section{\label{intro}Introduction}
One of the biggest goals in the field of high energy physics is to study the properties of QCD matter at various temperatures and baryon chemical potentials by exploring the QCD phase diagram. Experimentally this can be achieved by performing heavy-ion collisions at different energies. One way of studying the equation of state of QCD matter are fluctuations of conserved charges \cite{Asakawa:2015ybt}.
Fluctuations or more precisely cumulants are interesting as they can be related to the grand canonical partition function $Z^{\mathrm{GCE}}$ of the underlying theory. It has been pointed out that fluctuations of the net proton number are sensitive to a possible existence of a critical end point \cite{Stephanov:1999zu, Hatta:2003wn}.

From the theoretical side, lattice QCD calculations are the most fundamental calculations and allow to assess thermodynamic properties of QCD matter \cite{Bazavov:2017dus, Borsanyi:2018grb}. However, they are limited to low chemical potentials and therefore not suitable for studying fluctuations at large chemical potentials, where a possible critical point is expected \cite{Stephanov:2006zvm}. As a result, one is limited to effective models of QCD in the regions of high baryon densities.

On the experimental side, the beam energy scan (BES) programme is aimed at measuring various observables as a function of the colliding energy $\sqrt{s}$ \cite{Kumar:2013cqa} and measurements of fluctuations of the net proton number were recently published \cite{STAR:2020tga, STAR:2021iop}. Due to technical reasons only the net proton number can be measured in these experiments, though these are thought to serve as a proxy of the net baryon number \cite{Hatta:2003wn}. In comparison to the net baryon number however, the net proton number is not strictly conserved in QCD. In the future the Compressed Baryonic Matter (CBM) experiment will investigate fluctuation observables at even lower beam energies at FAIR (Facility for Antiproton and Ion Research). 

Charge conservation changes the cumulants because the fluctuations do not origin from an infinitely large heat bath as in the case of grand canonical ensemble \cite{Vovchenko:2021gas}. A way of limiting the effect of charge conservation in a heavy-ion collision is by minimizing the window of acceptance, in which the cumulants are measured \cite{Koch:2008ia}. On the other hand, when reducing it below the size of the correlations the fluctuations go to the Poissonian limit \cite{Bzdak:2017ltv}.

In \cite{Bzdak:2012an} analytic expressions of cumulants up to sixth order including perfect charge conservation were derived as a function of subvolume.
In \cite{Vovchenko:2020tsr}, a direct relation between between the grand-canonical susceptibilities of any theory and the fluctuations of a conserved charge in subvolumes of that theory has been derived.
These calculations however were performed within a static system and therefore no dynamical effects are accessible. Within \cite{Nahrgang:2009dqc}, a baseline calculation of fluctuations including conservation laws as a function of beam energy has been provided employing the UrQMD (Ultra-relativistic Quantum Molecular Dynamics) hadronic transport approach. Recently a model including critical dynamics has been employed to study the scaled variance \cite{Kuznietsov:2022pcn}. In \cite{Petersen:2015pcy} a calculation was performed using the SMASH (Simulating Many Accelerated Strongly-interacting Hadrons) approach within a box filled with hadronic matter. The effect of the net charge conservation was studied in a simplified interacting hadronic gas.

It is the purpose of this work to extend the calculation in \cite{Petersen:2015pcy} to the baryon number and more realistic interacting hadron gases. We use a hadronic transport approach (SMASH) and directly calculate fluctuations of conserved charges as a function of subvolumes of gases containing different sets of particles and interactions.
In addition, we want to address the question of the relation between the net proton and net baryon number fluctuations. A relation between the two quantities has been derived in \cite{Kitazawa:2012at} based on isospin diffusion in rapidity. These relations are important as in heavy-ion experiments only the net proton number fluctuations can be measured. Finally, the role of deuteron cluster formation is investigated and its impact on the net proton cumulants are calculated, as this has been pointed out as an important effect in \cite{Feckova:2015qza}. 

Throughout this work, fluctuations are calculated in coordinate space, even though experiments perform measurements in momentum space. At RHIC and LHC energies however, it is assumed that there is a strong correlation between momentum and coordinate space. For the sake of completeness, we extend our calculation to finite baryon chemical potential. 

The rest of the paper is organised as follows: First the main ingredients of the transport approach SMASH are introduced in Sec. \ref{smash}. Then in Sec. \ref{fluctuations} the methodology to extract the fluctuations of conserved charges is described. In Sec. \ref{simple_gas} the net charge fluctuations of a simplified system containing pions and rho mesons are analysed. In Sec. \ref{annihilation} the impact of a baryon annihilation is presented for a simplified hadron gas.
In Sec. \ref{FullSMASH} the baryon and proton number cumulants of the full SMASH hadron gas are presented. In Sec. \ref{mapping} the relation between the baryon and proton number fluctuation is discussed and finally in Sec. \ref{DeuteronFormation} the impact of deuteron cluster formation is shown in a simplified hadron gas.

\section{\label{smash}SMASH transport approach}
  The fluctuations of conserved charges and impact of hadronic interactions are calculated within the hadronic transport approach SMASH \cite{Weil:2016zrk, SMASH_github}. The specific version of the code which has been used is SMASH-2.0 \cite{dmytro_oliinychenko_2020_4336358}. Hadronic transport approaches are successful in describing the evolution of heavy-ion collisions at low beam energies or the late stages of ultra-relativistic heavy-ion collisions. SMASH has been applied to various collision energies \cite{Staudenmaier:2020xqr, Mohs:2019iee, Steinberg:2019wgm}, as well as it has been used to explore equilibrium properties of interacting hadronic matter by employing a box with periodic boundary conditions to simulate an infinite matter scenario \cite{Rose:2017bjz, Hammelmann:2018ath, Rose:2020sjv}. This is what we are interested in for this study concerning fluctuations. 
  
  At initialization, the particles are uniformly distributed inside the box. The momenta are sampled from the equilibrium Boltzmann distribution at a given temperature. 
  The incorporated interactions between the particles are resonance formations and binary scatterings. String excitations are not used within our calculations as they would break detailed balance. Resonances are modeled via relativistic Breit-Wigner distributions with a peak around the pole mass and widths depending on the mass \cite{Manley:1992yb}. The resonance lifetime is proportional to the inverse of the total width of the distribution. In the calculations presented here, no testparticles have been employed. 
  
  In this work, the geometric collision criterion is applied. It allows for a collision between two particles if the following relation $d_\perp < \sqrt{\sigma_{\rm tot} / \pi}$ between the transverse distance and the total cross section of the reaction is fulfilled. 
  SMASH contains all particles of the Particle Data Group up to masses of $\sim 2.3\,\rm GeV$ \cite{ParticleDataGroup:2020ssz}. Within the model, one can easily modify the particle content of the system and their interactions. We start therefore with a simplified system and study the impact of specific interactions before moving to the full set of particles, provided by SMASH.
  
\section{\label{fluctuations}Fluctuations in subvolumes}
  For our calculations, we employ a box with periodic boundary conditions representing infinite matter within SMASH. For each event, the box is initialized with the same number of particles. As a result, e.g. the net baryon number is conserved on an event-by-event basis. The initial number of each particle species is determined by taking one fully equilibrated event and plugging the resulting final multiplicities into each event, if not any other initial number of particles are specified. Besides thermal and chemical equilibrium, it is necessary that the density is distributed isotropically inside the box. Otherwise the result would depend on the definition of the subvolumes.
  
  \begin{figure}[h]
    \centering
    \includegraphics[width=0.48\textwidth]{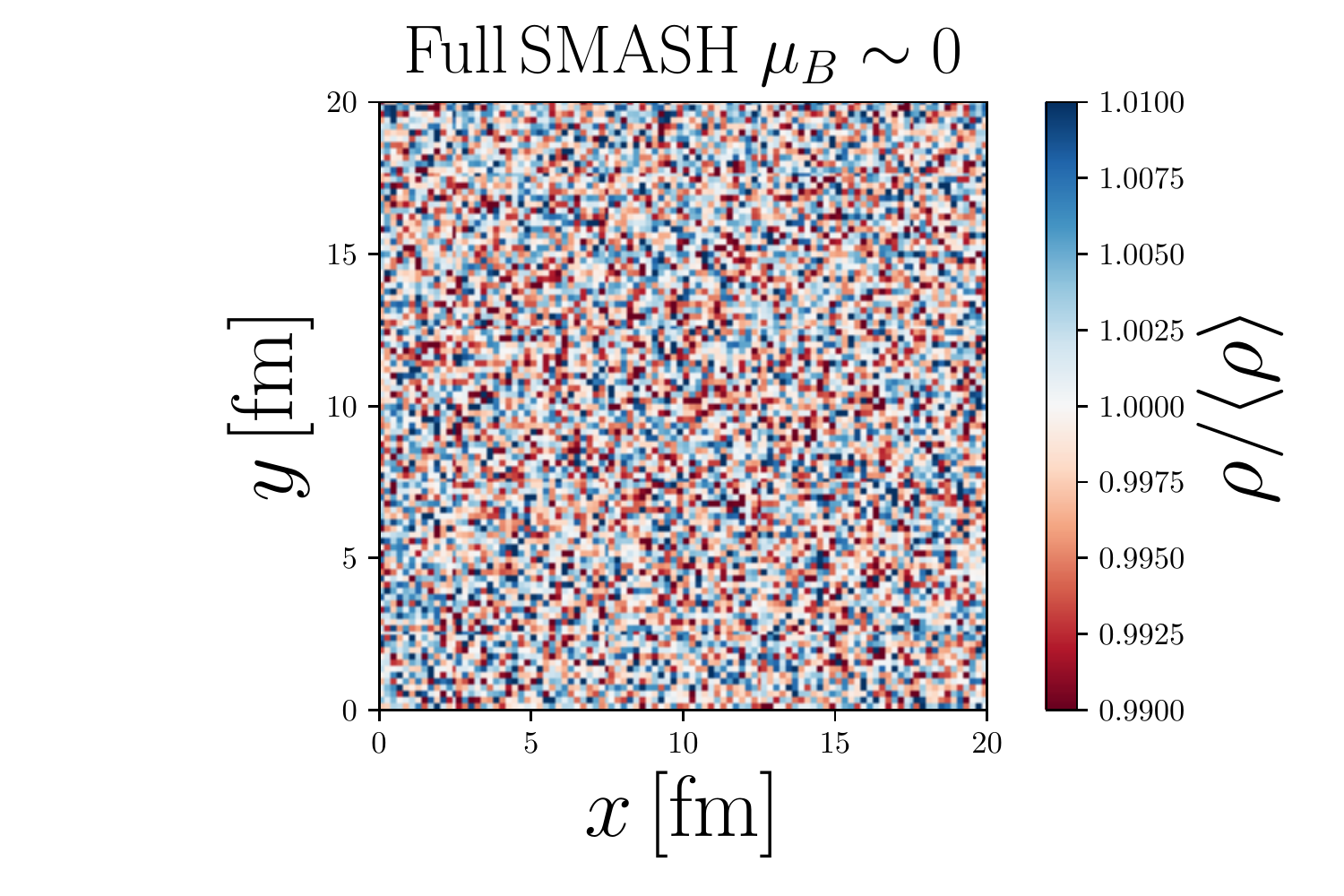}
    \caption{Normalized particle density in the $xy$-plane in the box containing the full set of particles in SMASH. The mean value of the hadron density is $\langle\rho\rangle = 0.253\,\mathrm{1 / fm^3}$. The volume of the box is $V = (20\,\mathrm{fm})^3$ and the density is computed after $t=150\,\mathrm{fm}$.}
    \label{Fig:HadronDensitynorm}
  \end{figure}
  
  \begin{figure}[h]
    \centering
    \includegraphics[width=0.45\textwidth]{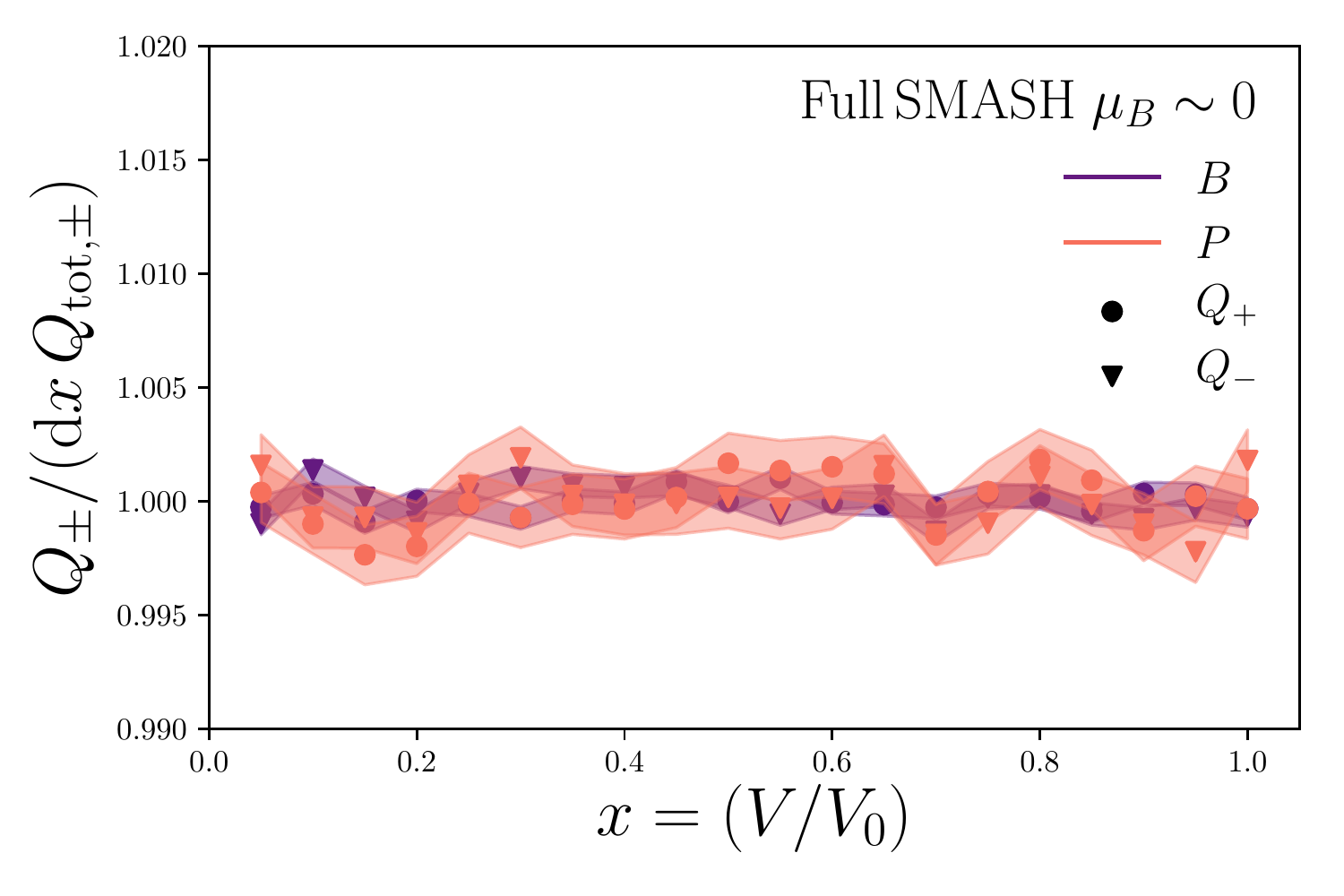}
    \caption{Positive and negative proton and baryon number distribution in each bin in the system with full SMASH hadron gas available. The proton number (orange) and baryon number (purple) are shown for both positive (circles) and negative (triangles) particles.}
    \label{Fig:ChargeDensity}
  \end{figure}
  
  Fig. \ref{Fig:HadronDensitynorm} shows the normalized hadron density inside the full SMASH box for $\mu_B\sim 0$. The density is shown in bins in the $xy$-plane. The $z$-axes is not taken into account. It shows that the density is distributed isotropically and no local spots of increased density appear. This a necessary pre-requisite to study the higher moments of the distributions. 
  
  Similarly Fig. \ref{Fig:ChargeDensity} shows in the case of the full SMASH hadron gas, the proton and baryon numbers are distributed equally in each bin.
  There exists a numerical artefact that has to be treated with care to avoid increased particle densities at the walls of the box. Within SMASH, interactions through the wall are searched for on a timestep basis, whereas collisions within the box are performed from action to action which in principle does not require a timestep. If one chooses too large timesteps, the code does not search for interactions through the wall and as a result more particles accumulate at the edges of the system and the density increases. To obtain reliable results with high statistics in this work, the choice of the timestep size has to be small enough to keep this artefact under control and the computing time on a reasonable level. It has been found that the timestep size should be $t_{\rm step} < 0.05\,\rm fm$.
 
\begin{figure}[h]
    \centering
    \includegraphics[width=0.5\textwidth]{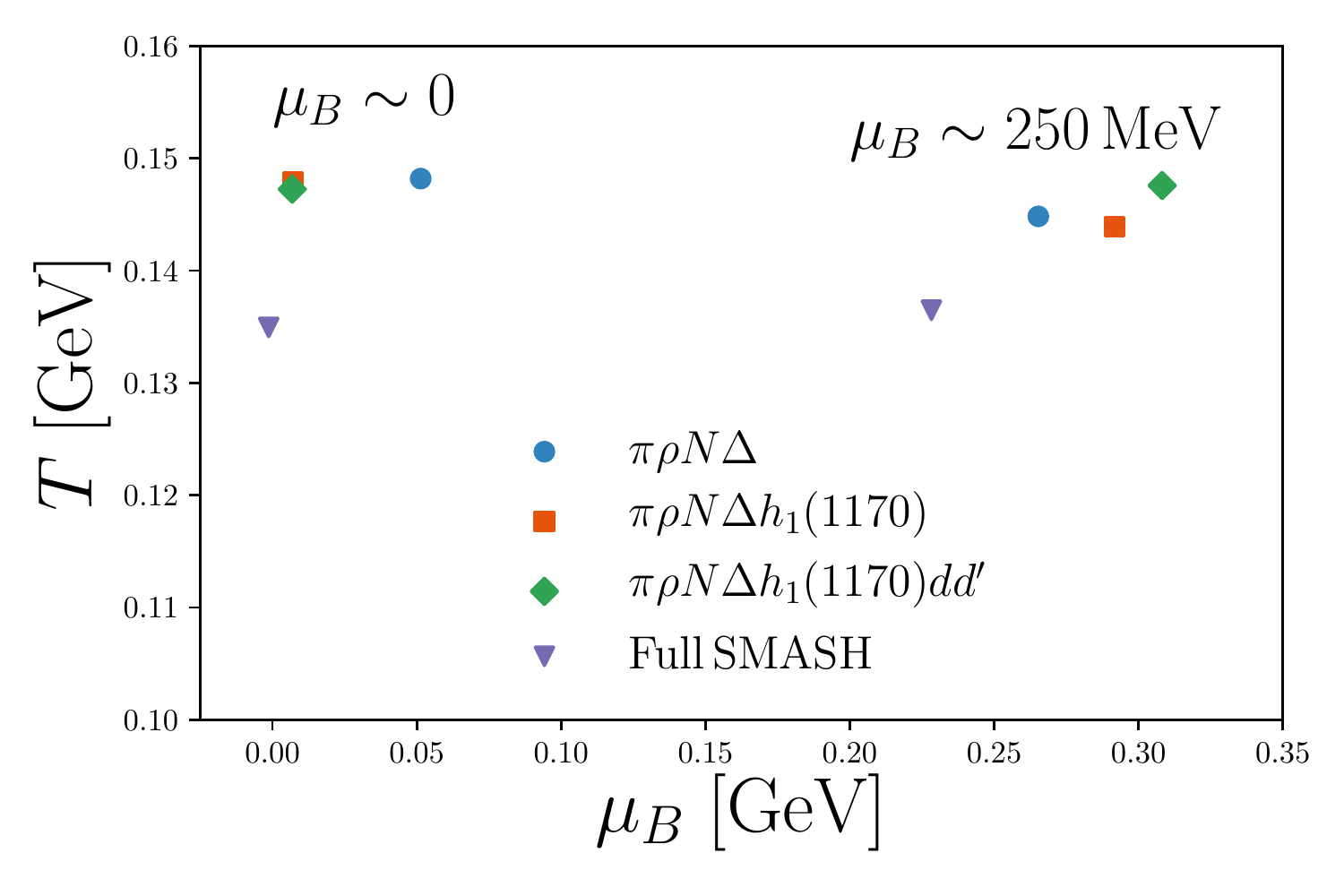}
    \caption{The final values of the temperature and baryon chemical potential for each system that is presented throughout this work.}
    \label{Fig:TmuBValues}
  \end{figure}

  After initializing the box with a given set of multiplicities, the temperature and baryon chemical potential are calculated in order to ensure, that the system has reached thermal and chemical equilibrium.
  Both $T$ and $\mu_B$ are calculated by assuming that the system follows the Boltzmann-statistics
  \begin{equation}\label{Eq:Boltzmann}
   \tfrac{dN}{dp} \sim e^{-\tfrac{\sqrt{m^2 + p^2} - \mu_B}{T}} \, .
  \end{equation} 
  In Fig. \ref{Fig:TmuBValues}, the temperature and baryon chemical potential are shown for each system after it has fully equilibrated. The boxes are initialized with the same temperature $T$ and baryon chemical potential $\mu_B$. At initialization, the systems is not in thermal and chemical equilibrium and as a result, inelastic scatterings, resonance formation and decays change the effective temperature. Especially in the full SMASH hadron gas interactions reduce the temperature compared to how it was initialized with. Even though there are deviations in the final temperatures of the systems, in \cite{Petersen:2015pcy} it was found that the cumulants show no large dependence on $T$.
  At initialization of the systems $T$ and $\mu_B$ were chosen from the freeze-out parametrization used in \cite{Bluhm:2016byc} for $\sqrt{s} = 2760\,\rm{GeV}$ and for $\sqrt{s} = 15\,\rm{GeV}$.
  The cumulants are presented in two groups, $\mu_B\sim 0$ and $\mu_B\sim 250\,\mathrm{MeV}$.
  
  After the box has equilibrated, subvolumes of a given size $V$ are defined. In each subvolume and per event, the net or total charge number $N_Q^{\rm net/tot} = N_{Q_+} \mp N_{Q_-}$ is obtained. Finally from each distribution the cumulants are computed as a function of the size of the subvolume $x = V / V_0$, where $V_0$ is the original volume of the system.
  The cumulants of e.g. the net baryon number are calculated as
  \begin{align}\label{Eq:CumulantRatios}
   C^{\mathrm{net}}_1 &= \langle N_B^{\rm net} \rangle \\
   C^{\mathrm{net}}_2 &= \langle (\delta N_B^{\rm net})^2 \rangle \\
   C^{\mathrm{net}}_3 &= \langle (\delta N_B^{\rm net})^3 \rangle \\
   C^{\mathrm{net}}_4 &= \langle (\delta N_B^{\rm net})^4 \rangle_c = \langle (\delta N_B^{\rm net})^4 \rangle - 3\langle (\delta N_B^{\rm net})^2 \rangle^2 \, .
  \end{align}
  Here, the brackets $\langle\cdot\rangle$ denote the sample average and $\delta N_i = N_i - \langle N_i\rangle$. Since the cumulants are proportional to the susceptibilities, one usually presents ratios in order to cancel additional factors in volume and temperature.
  Those are defined as
  \begin{align}
   \omega &= C_2^{\mathrm{net}} / C_1^{\mathrm{tot}} \\
   S\sigma &= C_3^{\mathrm{net}} / C_2^{\mathrm{net}} \\
   \kappa\sigma^2 &= C_4^{\mathrm{net}} / C_2^{\mathrm{net}} \, .
  \end{align}
  The errors of the cumulants are calculated according to \cite{Luo:2017faz}.
 
  Experimentally, only a limited set of particles (e.g. the net proton number) can be measured. In addition, restrictions in rapidity $\eta$ and transverse momentum $p_T$ \cite{Luo:2015ewa} limit the measurements even more.
  Since in SMASH, the full phase space information of each particle is available, we additionally study the impact of cuts in transverse momentum $p_T$ on the fluctuations of conserved charges by imposing
  \begin{align}\label{Eq:MomentumCut}
   0.4 &< p_T = \sqrt{p_x^2 + p_y^2} < 2 \,\mathrm{GeV} \, .
  \end{align}
  The value $x = (V / V_0)$ can be interpreted as the acceptance window \cite{Vovchenko:2021gas}. The number of events for all presented calculations are in the order of 5 million events.

\section{\label{simple_gas}Net electric charge fluctuations}
  To validate our methodology, the calculations in \cite{Petersen:2015pcy} are reproduced first. The considered system is the following: A box is filled with a fixed number of pions with initial momenta sampled according to the Boltzmann distribution at $T=160\,\rm MeV$.
  Their interaction is described by the formation of a $\rho$-meson ($\pi\pi\leftrightarrow\rho$) with an energy dependent cross section. After equilibrating the system both thermally and chemically, the fluctuations of the net charge number are calculated in the subvolumes of the box.
  
  In the case of only elastic interactions between the pions, the resulting fluctuations follow perfectly the equations derived in \cite{Bzdak:2012an}, which was also found in \cite{Petersen:2015pcy}. The analytic expressions are derived from taking two distinct Skellam distributions for each subvolume connected with a delta fixing the net charge number in the total volume (see Eq. 5 in \cite{Bzdak:2012an}).
  In the case of inelastic interactions, two different box volumes ($V = (10\,\mathrm{fm})^3$, $V = (20\,\mathrm{fm})^3$) and two different initial numbers of particles ($N_{start} = 100\pi^\pm$, $N_{start} = 200\pi^\pm$) were used.
  Since only equal numbers of positive and negative charged particles are incorporated, the odd number cumulants are zero and therefore $S\sigma$ is not shown.
  
  \begin{figure}[htp]
    \centering
    \includegraphics[width=0.99\linewidth]{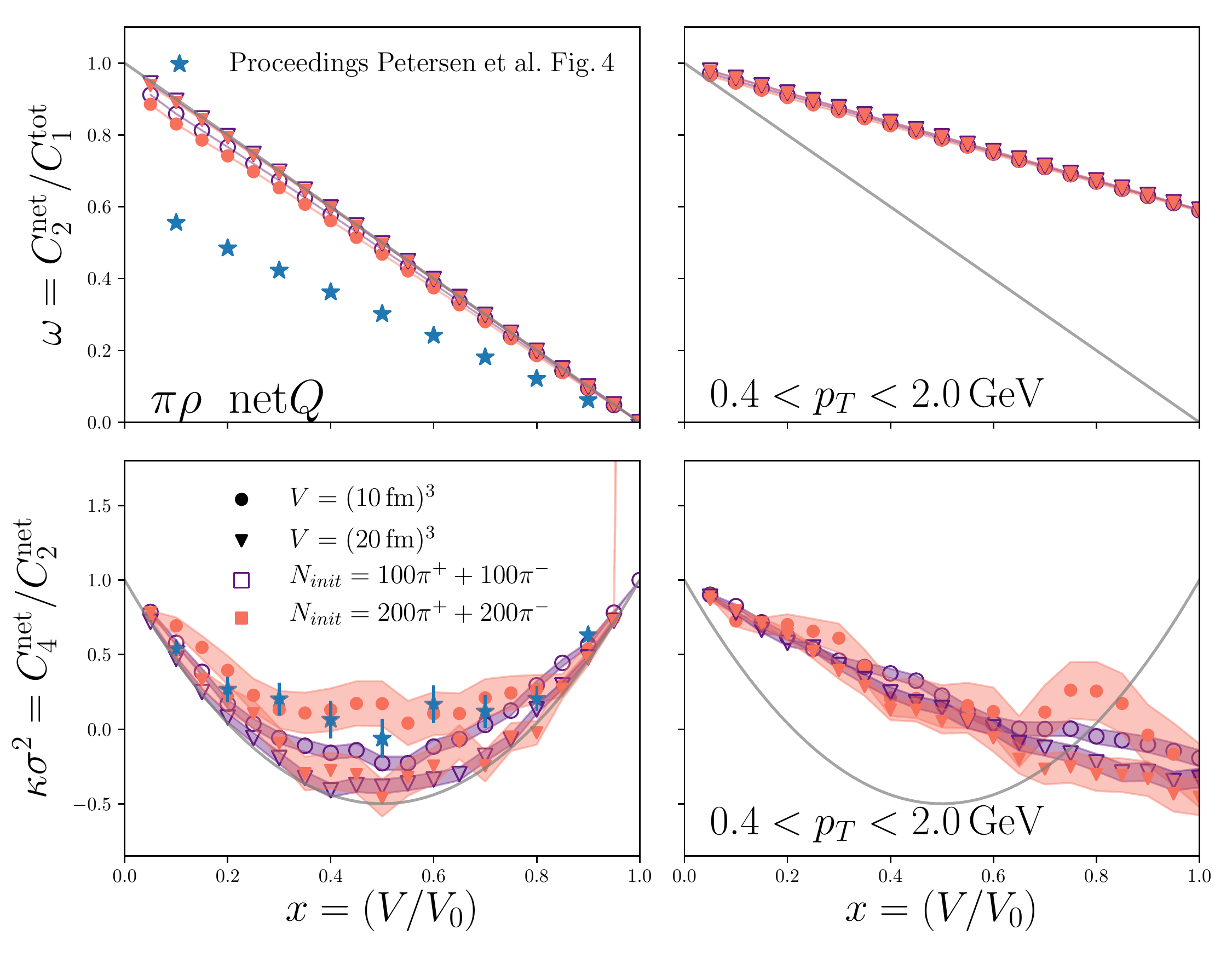}
    \caption{Scaled variance (top row) and kurtosis (bottom row) of the net charge distribution as function of the size of the subvolume. The full phase space (left column) and the restricted phase space (right column) are presented. The results are shown for the volumes $V=(10\,\mathrm{fm})^3$ (circles) and $V=(20\,\mathrm{fm})^3$ (triangles). The initial number of particles are $N_{\rm init} = 100\pi^+ + 100\pi^-$ (purple) and $N_{\rm init} = 200\pi^+ + 200\pi^-$ (orange). Additionally the results from \cite{Petersen:2015pcy} Fig.4 ($N_{\rm init} = 200\pi^+ + 200\pi^-$, $V = (10\,\rm{fm})^3$) for the full phase space are presented (blue stars).}  
    \label{Fig:PiRhoCumulantRatios}
  \end{figure}
  
  Fig. \ref{Fig:PiRhoCumulantRatios} shows the charge number cumulants as a function of the size of the subvolume. The scaled variance within the full phase space follows mainly the line $(1 - x)$. In the case of $V = (10\rm{fm})^3$ small deviations of the GCE value of $1$ for small values of $x$ appear. It has been checked that in the limit $x\rightarrow 0$ the scaled variance goes to $1$ which is referred to as the Poissonian limit.
  
  Contrary to \cite{Petersen:2015pcy} (see Fig. \ref{Fig:PiRhoCumulantRatios} blue stars) the scaled variance follows the conservation line even though the systems are equal. The difference between the current calculation and the prior work in \cite{Petersen:2015pcy} is the treatment of resonances in the last timestep. In the SMASH default calculations, all resonances decay into stable particles after the evolution. The decay products are then placed at the same position. This does not change the variance of the system, since the net charge is not affected if a positive and negative charged particle is at the same position. $C_1^{\rm tot}$ however is affected, since the total number of charged particles increases. Therefore $\omega$ no longer follows $1 - x$ but has a smaller slope. In this work, the $\rho$-meson is not forced to decay after the evolution.  
  When a momentum cut is imposed, the scaled variance is increased and no longer follows $1 - x$, see Fig. \ref{Fig:PiRhoCumulantRatios} upper right. $\omega$ still goes to 1 for $x\rightarrow 0$ but no longer goes to 0 for $x\rightarrow 1$. When only a subset of all particles is taken, the charge conservation only has a reduced effect on the cumulants.

  The kurtosis of the net charge number in full phase space is affected by the dynamical evolution and varies from the perfect conservation case $(1 - 6x(1-x))$. For $V=(10\rm{fm})^3$ and $N_{init} = 200\pi^+ + 200\pi^-$ the kurtosis is shifted towards $\kappa\sigma^2\sim 0$ around $x = 0.5$. For $N_{init} = 100\pi^+ + 100\pi^-$ and $V=(20\rm{fm})^3$ it is close to the perfect conservation case. The shape of $\kappa\sigma^2$ also matches with the calculation in \cite{Petersen:2015pcy}.
  The shift of the kurtosis can be reproduced in a simplified model, see Appendix \ref{toymodel}. Our finding is that when a large fraction of charged particles form a resonance with charge 0, the ratio $\kappa\sigma^2\rightarrow 0.25$ around $x=0.5$ where the conservation effect is the largest.
  In the case of an applied momentum cut, all four curves follow a similar trend towards $\kappa\sigma^2 = -0.5$ for $x\rightarrow 1$. Within errors, no clear distinction between the four cases can be made. For small subvolumes the slope with which the kurtosis approaches the GCE value differs from the kurtosis within the full phase space. The qualitative behaviour of the kurtosis when employing a momentum space cut is consistent with \cite{Karsch:2015zna}. 
  
  \begin{figure}[htp]
    \centering
    \includegraphics[width=1.0\linewidth]{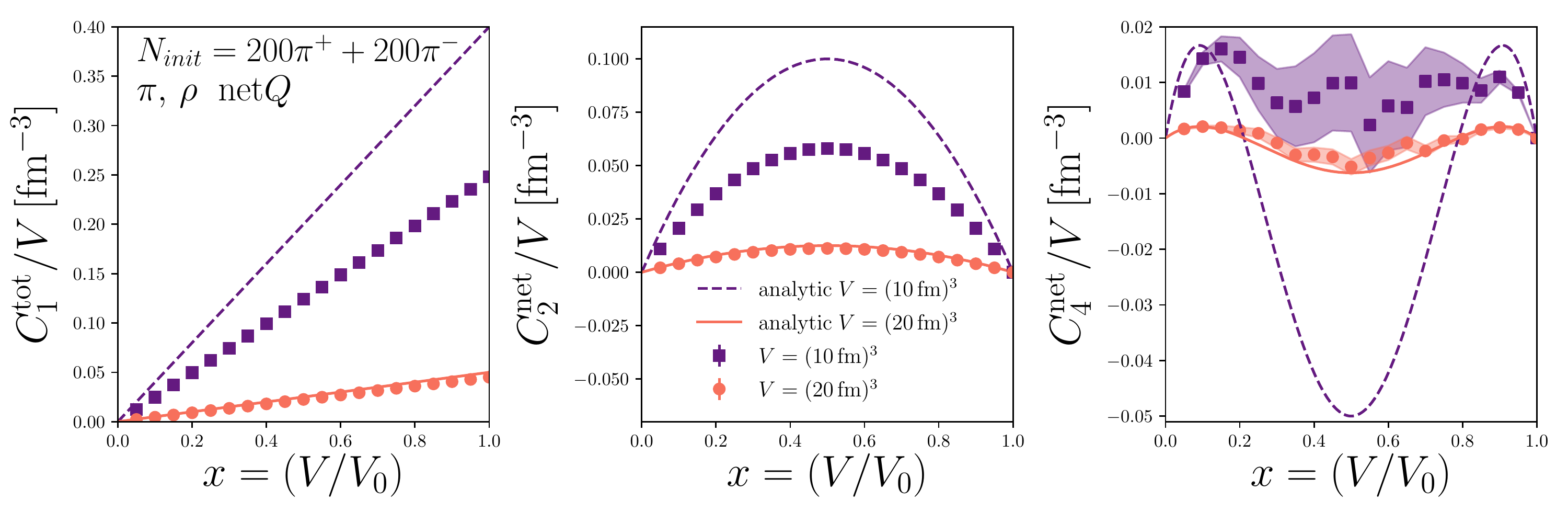}
    \caption{Cumulants of the total charge number $C_1$ (left), second cumulant $C_2$ (center) and the fourth cumulant $C_4$ (right) of the net charge number divided by the volume of the system as a function of the size of the subvolume. The systems are initialized with the same number of pions $N_{\rm init} = 200\pi^+ + 200\pi^-$. The results are presented for two distinct volumes $V=(10\,\mathrm{fm})^3$ (purple) and $V=(20\,\mathrm{fm})^3$ (orange). The lines show analytic expectation of each cumulant and the points show the results from SMASH after dynamically evolving of the box.}  
    \label{Fig:PiRhoC1C2C4}
  \end{figure}
  
A more detailed view of this effect is shown in Fig. \ref{Fig:PiRhoC1C2C4}, where the individual cumulants divided by the volumes are presented. The results are shown for two distinct cases $V=(10\,\rm{fm})^3$ and $V=(20\,\rm{fm})^3$ with the same initial number of pions $N_{init} = 200\pi^+ 200\pi^-$.
In both cases we find perfect agreement with \cite{Bzdak:2012an} for the cumulants at initialization, when there are no correlations. We also find good agreement after dynamically evolving the system in $V=(20\,\rm{fm})^3$. In the case of $V=(10\,\rm{fm})^3$ however the discrepancy between analytic expectation and numerical result is larger.
The main difference is the charge density (see Fig. \ref{Fig:PiRhoC1C2C4} left) and therefore the effective number of interactions of the type $\pi^\pm\pi^\mp\leftrightarrow\rho^0$, which annihilates a positive and negative charge. For $V=(10\,\rm{fm})^3$ we find a charge density of $\rho\sim 0.25\,\mathrm{fm^{-3}}$ and $N_{\rm{interactions}}\sim 80\,\mathrm{fm^{-1}}$. For $V=(20\,\rm{fm})^3$ the charge density is $\rho\sim 0.05 \,\mathrm{fm^{-3}}$ with $\sim 21\,\mathrm{fm^{-1}}$ interactions per unit time.
As a result of the increased number of interactions and neutral charged particles ($\rho^0$), the variance $C_2^{\mathrm{net}}$ of the net charge distribution is reduced and the fourth cumulant is shifted towards positive values around $x=0.5$, whereas the analytic curve goes to negative values. This is the reason for the ratio $\kappa\sigma^2$ (Fig. \ref{Fig:PiRhoCumulantRatios} lower left) being positive.

We finally conclude that the density and interaction rate of producing charge 0 particles plays an important role when studying cumulants of the net charge number. Additionally we find good agreement with \cite{Bzdak:2012an} for a rather dilute system. In a very dense system, the cumulants differ more strongly from the derived expressions even though the net charge number is perfectly conserved.

\section{\label{NNbar}Net baryon number fluctuations}
In this section, the influence of net baryon number conservation on the cumulants is studied. As a start, the impact of a baryon annihilation process in a simplified hadronic system is investigated. At higher beam energies annihilation processes are important in the late stage hadronic rescattering phase \cite{Garcia-Montero:2021haa, Savchuk:2021aog}. Since the geometric collision criterion is employed the representative baryon annihilation process $B\bar B\leftrightarrow 5\pi$ has to be modeled with binary interactions. In SMASH default calculations, annihilation processes are performed via string excitations, but as they break detailed balance this formalism cannot be used in the box.
Another option would have been to employ a recently implemented stochastic collision criterion, which allows for multi-particle interactions \cite{Staudenmaier:2021lrg, Garcia-Montero:2021haa}. However, as this more sophisticated multi-particle treatment increases the runtime of the code, we stick to the geometric criterion in this work.
With only binary scatterings, the nucleon-nucleon annihilation process is modeled via an intermediate resonance formation and decay. The corresponding processes, which results in an effective $N\bar N\rightarrow 5\pi$ reaction are
\begin{flalign}\label{Eq:NNbarAnnihilation}
  &N\bar N \leftrightarrow h_1(1170)\rho&
  &h_1(1170) \leftrightarrow \rho\pi&
  &\rho \leftrightarrow \pi\pi \, .
\end{flalign}
as suggested in \cite{Demir:2008tr}. 

\subsection*{\label{annihilation}Baryon annihilation}
  Let us now quantify the influence of baryon annihilation on the baryon number cumulants. For this purpose the fluctuations of a simplified hadron gas with and without an annihilation process are calculated (see system 2 and 3 in Appendix \ref{App:Systems}). The process $N\bar N\leftrightarrow 5\pi$ is performed via the intermediate resonance process $N\bar N\leftrightarrow h_1(1170)\rho\leftrightarrow 5\pi$.
  System 2 contains the same baryon species, however the interactions are chosen such that only the $2\leftrightarrow 1$ reaction happens ($\Delta\leftrightarrow N\pi$). As the interaction between $N$, $\pi$ and $\Delta$ does not alter the baryon number, as consequence and contrary to system 3, system 2 not only conserves the net baryon number, but also the total baryon number. 
  
  \begin{figure}[htp]
    \centering
    \includegraphics[width=0.99\linewidth]{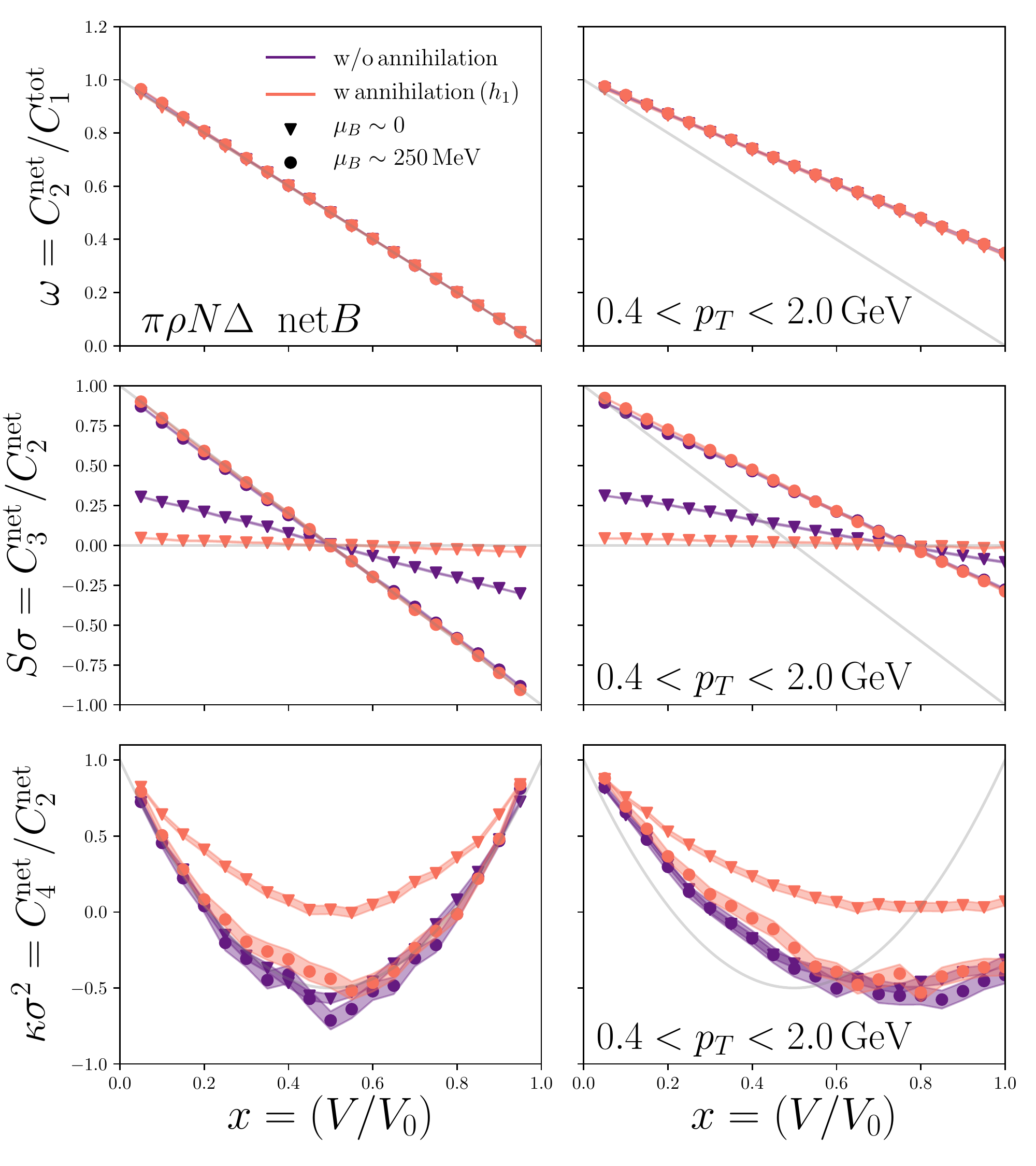}
    \caption{Scaled variance (top), skewness (center) and kurtosis (bottom) as a function of the size of the subvolume. The blue points correspond to system 2 without $N\bar N$ annihilation and the yellow points to system 3 with the $N\bar N$ annihilation process. Both calculations are shown for $\mu_B\sim 0$ (triangles) and for $\mu_B\sim 250\,\mathrm{MeV}$ (circles).} 
    \label{Fig:W_WO_annihilation}
  \end{figure}
  
  The result for the baryon number fluctuations of both systems are shown in Fig. \ref{Fig:W_WO_annihilation} for two different values of baryon chemical potentials. Both results are calculated in a box with $V=(15\,\rm{fm})^3$. Since we want to focus on the influence of the annihilation process on the fluctuations, forced decays into stable particles after the final time step are not performed here.
  The scaled variance is the same for both systems and independent of the chemical potential or type of interactions following the line $(1 - x)$. With an additional cut in momentum space, $\omega$ is still influenced by the net baryon conservation. But due to the phase-space limitations, it is not fully conserved anymore and does not reach 0 for the full volume.
  
  The skewness is influenced by the baryon chemical potential, since it is sensitive to asymmetries of charges to anti charges. In the limit of large $\mu_B$ it follows $(1 - 2x)$, whereas in the limit of small $\mu_B$ it is approximately zero. Because system 2 evolves into a small but non-zero value of baryon chemical potential within the dynamical evolution ($\mu_B\sim 0.05\,\rm{GeV}$, see Fig. \ref{Fig:TmuBValues} blue circles) the skewness also shows a small non-zero slope. However there is no clear influence of a dynamical baryon annihilation on $S\sigma$ visible.
  With an applied $p_T$ cut and similar to the scaled variance, the cumulants are not as strongly affected by the baryon number conservation and follow a reduced slope towards to full volume $x=1$.
  
  Similar to what we have observed in Sec. \ref{simple_gas}, the kurtosis is strongly influenced by the dynamical evolution of the system, when a baryon annihilation process in form of the $h_1(1170)$ meson is added. Here two can effects can be seen. First for $\mu_B\sim 0$ the kurtosis is shifted towards positive values and secondly, $\kappa\sigma^2$ goes to $-0.5$ at $x=0.5$ for large values of baryon chemical potential. This has been found by \cite{Vovchenko:2020tsr}, where it was calculated that the kurtosis is additionally influenced by $C_3 / C_2$ (see Eq. 18 in \cite{Vovchenko:2020tsr}). We find that this effect is driven by the annihilation process. When a large portion of the nucleons and antinucleons have formed pions, the kurtosis is affected. At large values of baryon chemical potential there are not enough antinucleons in the system to perform the reaction in the first place, therefore the kurtosis is not affected.
  With an additional cut on the transverse momentum, there is still a clear distinction between the cases with and without baryon annihilation. This means that even when only a reduced set of particles is taken into account, where the effect of the baryon number conservation is not as strong anymore, the process $N\bar N\rightarrow 5\pi$ has a strong influence on the kurtosis. Compared to \cite{Savchuk:2021aog} we only see an effect of the annihilation process on the fourth cumulant.

\subsection*{\label{FullSMASH}Full SMASH calculation}
In this section the result of the full SMASH hadron gas with all of its interactions is presented (for more details see \cite{SMASH_github}). For this purpose a box with $V=(20\,\rm{fm})^3$ is employed. Experimentally only stable particles are measured since all resonances decay into ground states. Therefore for the results in this Section, after the final timestep all resonances are forced to decay until all decay products are stable. For technical reasons the decay products are placed at the same position as the original resonance.
This affects only the proton number cumulants, since more protons appear from the decay chain. The baryon number cumulants are not affected, because all decays in SMASH conserve the baryon number.
The cumulants therefore can be viewed as the sum of dynamical part and final decay part $C_i^{\rm{net\, P}} = C_i^{\rm{net\, P,\, dynamic}} + C_i^{\rm{net\, P,\, decays}}$. In the previous section a focus has been set on the baryon annihilation process. In this section, the focus will be on the proton number cumulants, since they are the ones that are actually measured in experiment.

\begin{figure}[htp]
  \centering
  \includegraphics[width=0.99\linewidth]{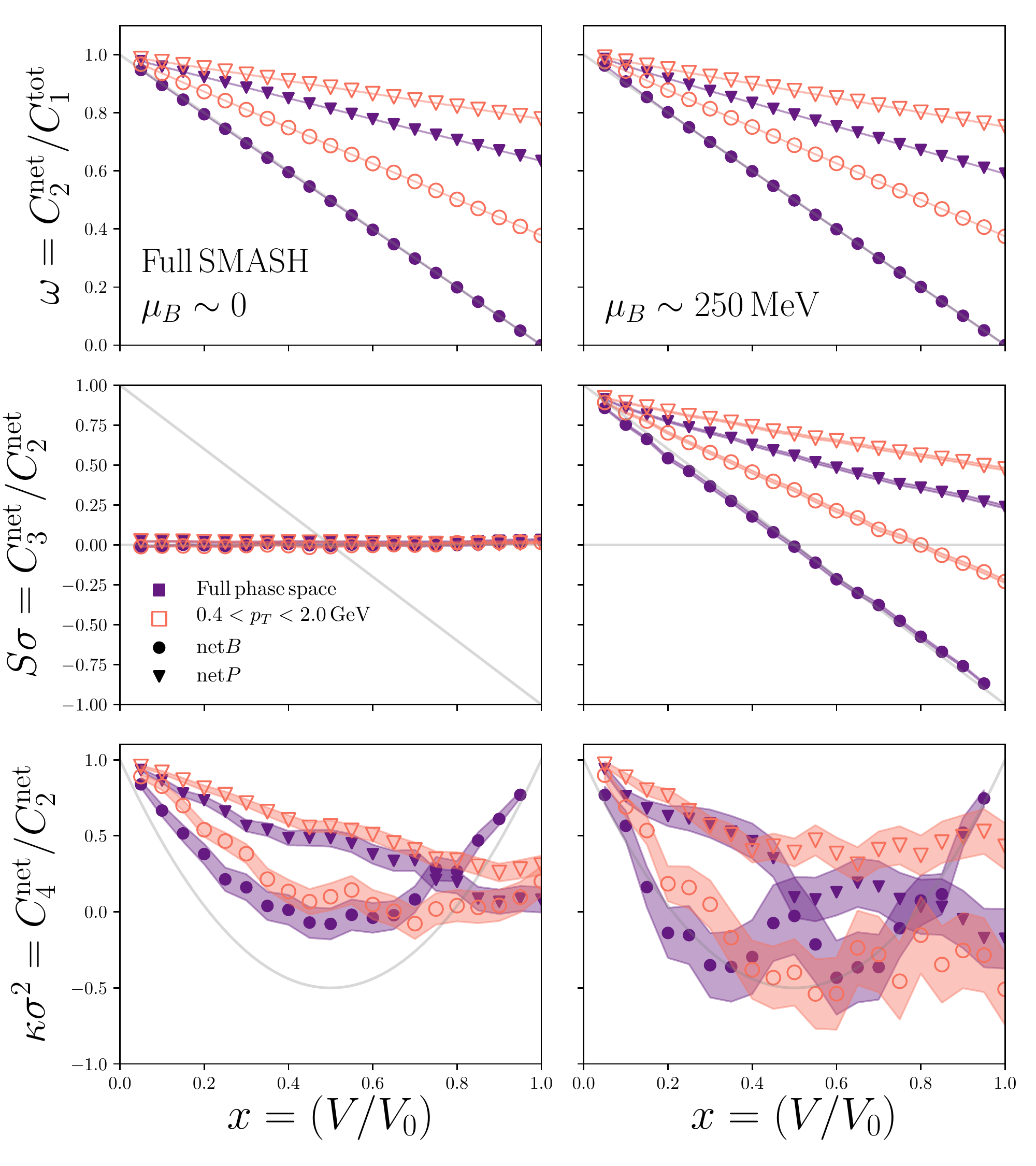} 
  \caption{Scaled variance (top), skewness (center) and kurtosis (bottom) as function of the size of the subvolume of the full SMASH hadron gas. The results of the baryon number (circles) and proton number cumulants (triangles) are shown for $\mu_B\sim 0$ (left) and $\mu_B\sim 250\,\rm MeV$ (right).
           The cumulants are presented for the full phase space (purple) and restricted phase space (orange).}
  \label{FullSMASHCumulants}
\end{figure}

Fig. \ref{FullSMASHCumulants} shows the result of baryon and proton number cumulants of the full SMASH hadron gas. The baryon number cumulants (circles) in full phase space follow the conservation lines and are similarly influenced by the annihilation process as the simplified hadron gas discussed in the previous section.
However for $\mu_B\sim 250\,\rm MeV$ the kurtosis has a much larger error than in the case of $\mu_B\sim 0$, which is why it is hard to draw conclusions. There seems to be the trend observed in the previous Section that the annihilation process plays no dominant role on the cumulants at large baryon chemical potentials in the full SMASH hadron gas. The net baryon number within a restricted phase space is again not conserved anymore. However they are still affected by the baryon number conservation and are not completely thermal.

Similarly to the net baryon number in restricted momentum space, the net proton number is not conserved and therefore the cumulants do not follow the conservation curves. Nevertheless they are affected by the net baryon number conservation. The scaled variance of the proton number cumulants for both $\mu_B\sim 0$ and $\mu_B\sim 250\,\rm MeV$ have a negative slope in the full and restricted phase space. With $p_T$ cut, the slopes are even more reduced, since an even smaller subset of the conserved charge is taken into account.
The slopes of the proton scaled variance are shown in Tab. \ref{TableFitsOmega}.

\begin{table}[hbt!]
  \centering
  \begin{tabular}{l | c c}
   $\omega(x) = C_2^{\rm{net}P} / C_1^{\rm{tot}P}$ & $\mu_B\sim 0$ & $\mu_B\sim 250\,\rm{MeV}$ \\\hline
   full phase space & $0.995 - 0.361x$ & $1.001 - 0.412x$ \\
   $0.4 < p_T < 2.0\,\rm{GeV}$ & $0.998 - 0.219x$ & $1.001 - 0.240x$ \\
  \end{tabular}
  \caption{Fits of the proton scaled variance as a function of the size of the subvolume $x$}
  \label{TableFitsOmega}
\end{table}

The values of the slope of the net proton scaled variance in both full and restricted momentum space indicate that they are also influenced by the annihilation process as they differ between $\mu_B\sim 0$ and $\mu_B\sim 250\,\rm{MeV}$. Additionally, the cumulants as a function of the size of the subvolume of the net proton number deviate from the perfect conservation case. As a result they will not follow these lines in an experimental situation and the differences between the net protons and net baryons has to be taken into account.

\section{\label{mapping}Net baryon vs net proton fluctuations}
  In this section, the relation between the net proton and the net baryon number fluctuation is investigated. For this purpose, the expressions derived in \cite{Kitazawa:2012at} are tested.
  Since experiments only have access to the net proton number and not the full baryon number spectrum, the net proton number is used as a proxy of the net baryon number.
  In most theoretical calculations however, the net baryon number is calculated and the net proton number is not accessible. Therefore it is important to know if a relation between the two quantities exists. 

  It is argued by the authors of \cite{Kitazawa:2012at} that the mapping works if the underlying system undergoes isospin randomization. This process happens e.g. through the $\Delta$ resonance and a pion at large collision energies where many pions are produced. At small collision energies, isospin randomization is expected to break down, since there are not enough pions produced. Generally, the mapping is derived via a binomial factorization ansatz which maps a proton onto a baryon with a certain probability.
  The respective formulas for the first two cumulants are 
  \begin{align}\label{Eq:Mapping}
    \left\langle N_B^{\rm{tot}} \right\rangle = \left\langle \xi_1^{-1} N_p + \bar{\xi}_1^{-1} N_{\bar p} \right\rangle \\
    \left\langle \left( \delta N_B^{\rm{net}}\right)^2 \right\rangle = \left\langle \left(\xi_1^{-1}\delta N_p - \bar{\xi}_1^{-1}\delta N_{\bar p} \right)^2 \right\rangle \\\nonumber
                                                                     -   \left\langle \xi_2\xi_1^{-3} \delta N_p + \bar{\xi}_2\bar{\xi}_1^{-3} \delta N_{\bar p} \right\rangle \, .
  \end{align}
  Here $\xi_1 = p$ and $\xi_2 = p(1-p)$ with the probability $p = \langle N_p \rangle / \langle N_B \rangle$ and respective for the anti particles $\bar \xi$. To have an estimate of an error on the result, the probability is modified by $\pm 3\%$ and presented in bands.
  We have checked that for both small and large baryon chemical potentials the isospin density $\alpha = \tfrac{1}{2}\left(\frac{\langle N_n\rangle - \langle N_p\rangle}{\langle N_n \rangle + \langle N_p\rangle}\right) \ll 1$. 
  \begin{figure}[htp]
    \centering
    \includegraphics[width=0.99\linewidth]{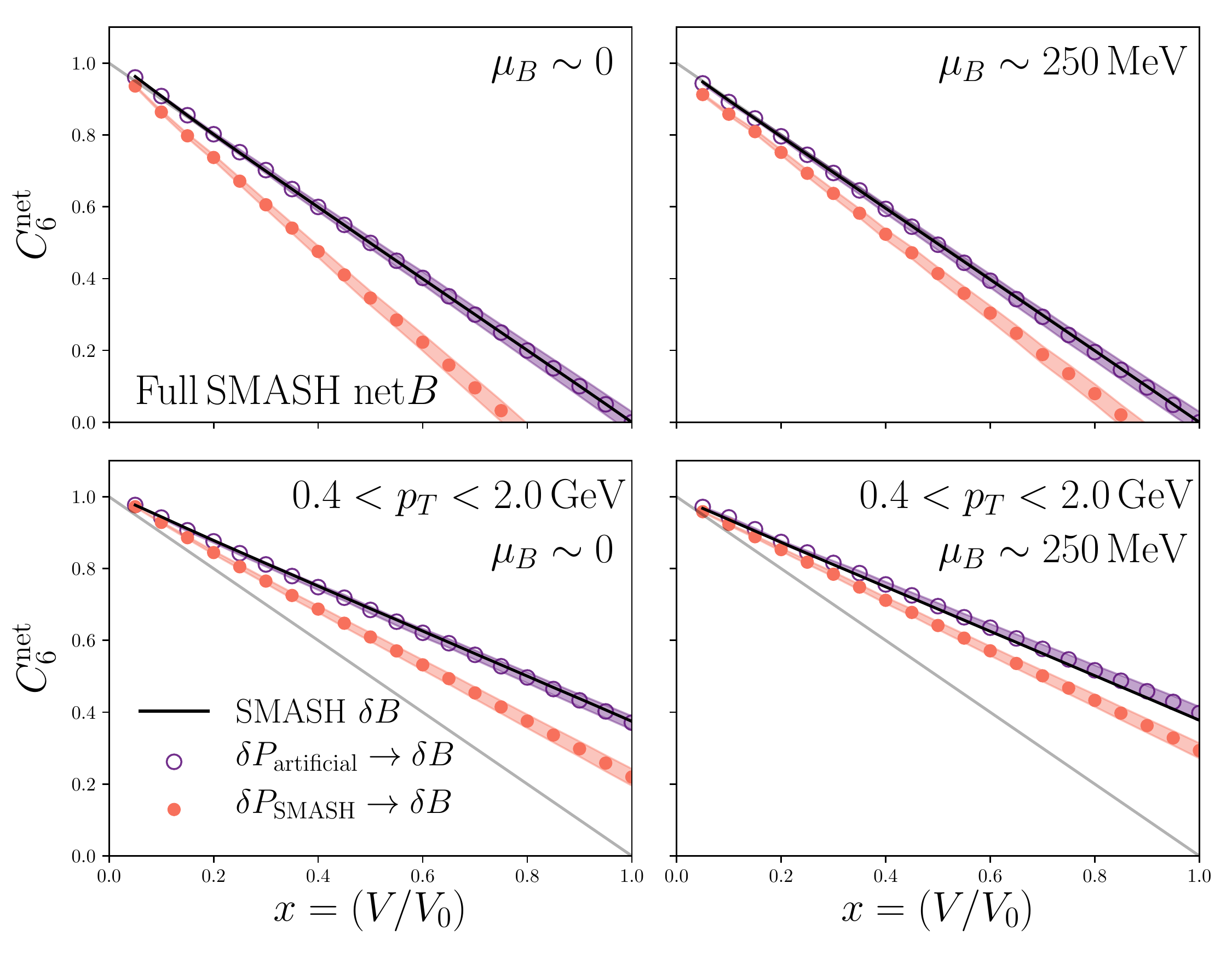} 
    \caption{Scaled variance as function of the size of the subvolume. The result are shown for $\mu_B = 0$ (left column) and $\mu_B\sim 250\,\rm MeV$ (right column).
             Fluctuations of the full phase-space are shown on the top row and restricted momentum space on the bottom row. The grey line is the analytic expectation of perfect conservation.
             The black line is the result of the actual net baryon scaled variance. The orange points are the result if one applies the mapping \cite{Kitazawa:2012at} onto the net proton number.
             The purple points show the result when mapping the fluctuations of an artificial set of protons back onto the baryons. 
             }
    \label{FigMappingC2}
  \end{figure}
  
  Fig. \ref{FigMappingC2} shows the result of mapping the net proton number onto the net baryon number in the full SMASH hadron gas after dynamically evolving the system and performing final decays into all ground state particles. As the net baryon number is conserved it follows $(1 - x)$. To assess how well the mapping proposed in \cite{Kitazawa:2012at} works, let us compare two different situations. The purple points are obtained by generating an artificial set of protons by selecting event-by-event a (anti-) baryon as a (anti-) proton with the given probability $p\,(\bar p) $. The orange circles display the actual set of protons in the SMASH calculation. Afterwards, Eqns. \ref{Eq:Mapping} are applied with the same probabilities to get the net baryon cumulants. As shown in Fig. \ref{FigMappingC2}, when starting with the artificial proton set, the mapping can reconstruct the scaled variance of the net baryon number fluctuations. 
  
  However when applying the mapping on the actual SMASH net proton fluctuations, the scaled variance of the net baryon number cannot be fully reconstructed.
  At small values of $x$ the differences are not large. At large $x$ when the conservation of the net baryon number becomes important, the fluctuations are underestimated.
  Even in restricted momentum space these correlations are still present as the scaled variance can still not be reconstructed.
  It has been checked that this difference comes mainly from the variance $C_2^{\rm net}$ and not from the mean $C_1^{\rm tot}$, which is perfectly reproduced.
  This is a result of dynamical correlations within the set of protons. As in the case of the artificial proton set, all correlations are removed by the binomial acceptance.
  
  \begin{figure}[htp]
    \centering
    \includegraphics[width=0.99\linewidth]{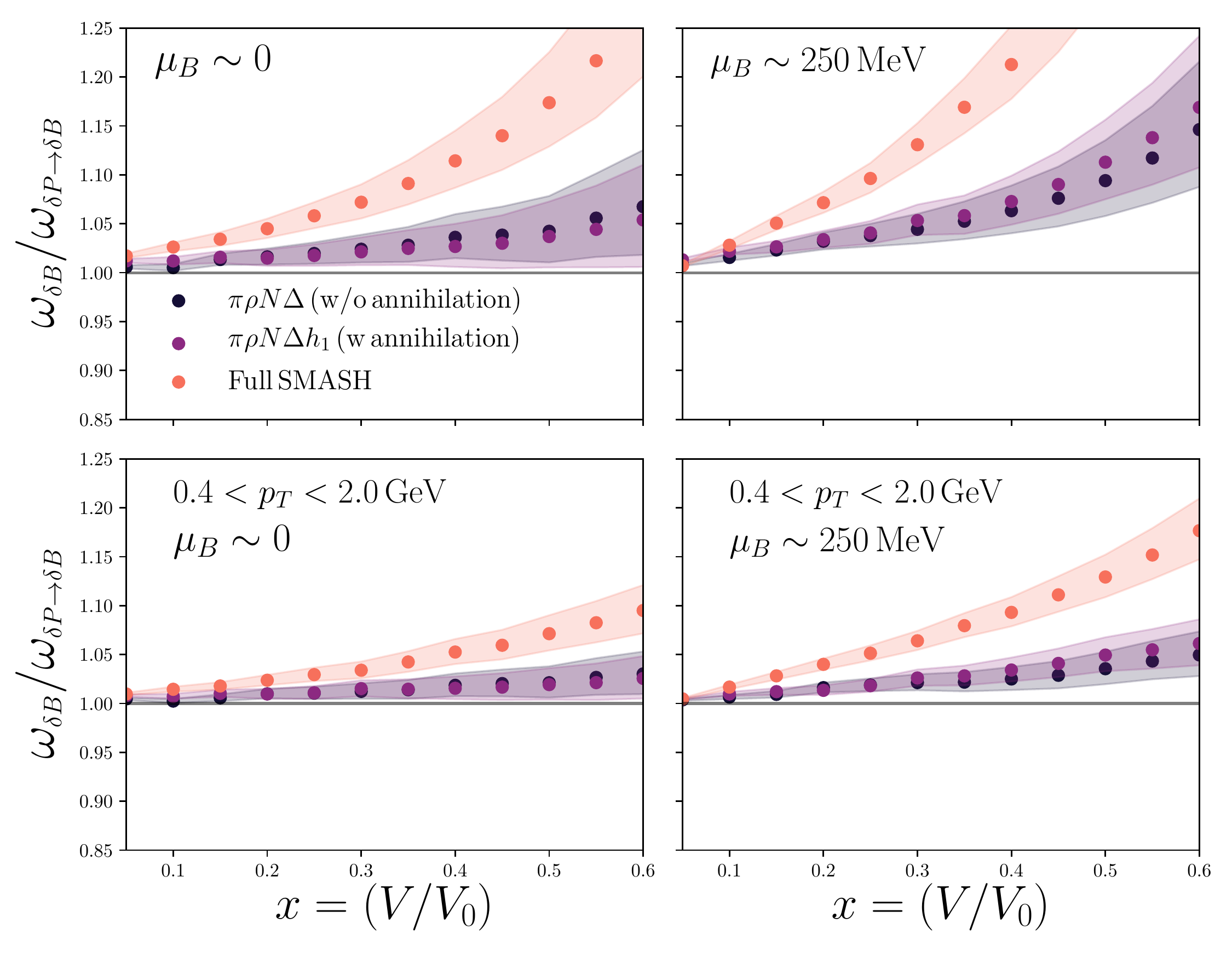} 
    \caption{Ratio of the scaled variance of baryons and baryons when mapped from the protons $\omega_{\delta B} / \omega_{\delta P\rightarrow\delta B}$ as a function of the size of the subvolume.
    Three different systems with increasing number of degrees of freedom and interactions are shown. In black a system containing $\pi\rho N\Delta$ is used, where no annihilation process is implemented.
    In purple, the same system is used but with an annihilation process and in orange, the full SMASH hadron gas is used.}
    \label{FigMappingC2Ratios}
  \end{figure}
  Fig. \ref{FigMappingC2Ratios} shows the ratio of the scaled variance of baryons and protons when mapped onto the baryons for three different systems. For all three systems the difference grows with increasing subvolume size where the effect of conservation is the largest.
  In the case of the full SMASH hadron gas, the difference becomes the largest as there are more dynamical correlations within the system. From the two simplified systems one can see that the baryon annihilation process is not the reason why the binomial unfolding cannot reconstruct the scaled variance of the baryons.
  We also calculated the mapping $\delta B\rightarrow\delta P$ and found that the scaled variance $\omega^{\rm{net}P}$ could be perfectly reconstructed from the baryon number. Our findings can be summarized in the following statement
  \begin{equation}
  \delta P^{\mathrm{SMASH}}_{\mathrm{not\, conserved}} \xrightleftharpoons[\checkmark]{\times} \delta B^{\mathrm{SMASH}}_{\mathrm{conserved}} \, ,
  \end{equation}
  meaning the proton cumulants can be reconstructed from the baryons but not vice versa.
  
\section{\label{DeuteronFormation}Deuteron formation}
In this section, the influence of deuteron cluster formation on conservation effects is calculated. For this, two different sets of particles and interactions are employed, where the only difference is the effective deuteron formation process (see Eq. \ref{Eq:DeuteronFormation}).
The deuteron cluster formation is an important process when studying fluctuations since they are produced in the late stages of a heavy-ion collision. An analysis of the influence of deuteron cluster formation on the net proton number fluctuations can be found e.g. in \cite{Feckova:2016kjx}.

In this work we want to determine the effect of deuteron formation on conservation effects of the proton and baryon number cumulants. By comparing systems with and without a deuteron cluster formation process the impact on the proton number cumulant can be studied as a function of the size of the subvolume.
With a geometric collision criterion, limitations are that only binary scatterings can be performed. The reaction in which a deuteron is created is a $3\leftrightarrow 2$ reaction namely $\pi n p \leftrightarrow d\pi$ and $N n p \leftrightarrow Nd$.
To be able to perform these interactions a fictional particle $d^\prime$ is introduced \cite{Oliinychenko:2018ugs,Oliinychenko:2020znl}. Note that the deuteron in this microscopic description is treated as a point particle. The individual reactions that model the $3\leftrightarrow 2$ interaction $\pi n p \leftrightarrow d\pi$ and $N n p \leftrightarrow Nd$ are 
\begin{flalign}\label{Eq:DeuteronFormation}
  &pn \leftrightarrow d^\prime&
  &\pi d^\prime \leftrightarrow \pi d&
  &N d^\prime \leftrightarrow N d \, .
\end{flalign}

We now study the impact of deuteron clusters on conservation curves of the proton and baryon cumulants. To do this the cumulants as a function of $x=(V/V_0)$ are calculated for a system with and without deuterons (see system 3 and 4 in Appendix \ref{App:Systems}) in a box of $V=(15\,\rm{fm})^3$.
For the calculation of cumulants of deuterons, only the actual deuterons are counted and not the fictional $d^\prime$ resonance.
\begin{figure}[htp]
  \centering
  \includegraphics[width=0.99\linewidth]{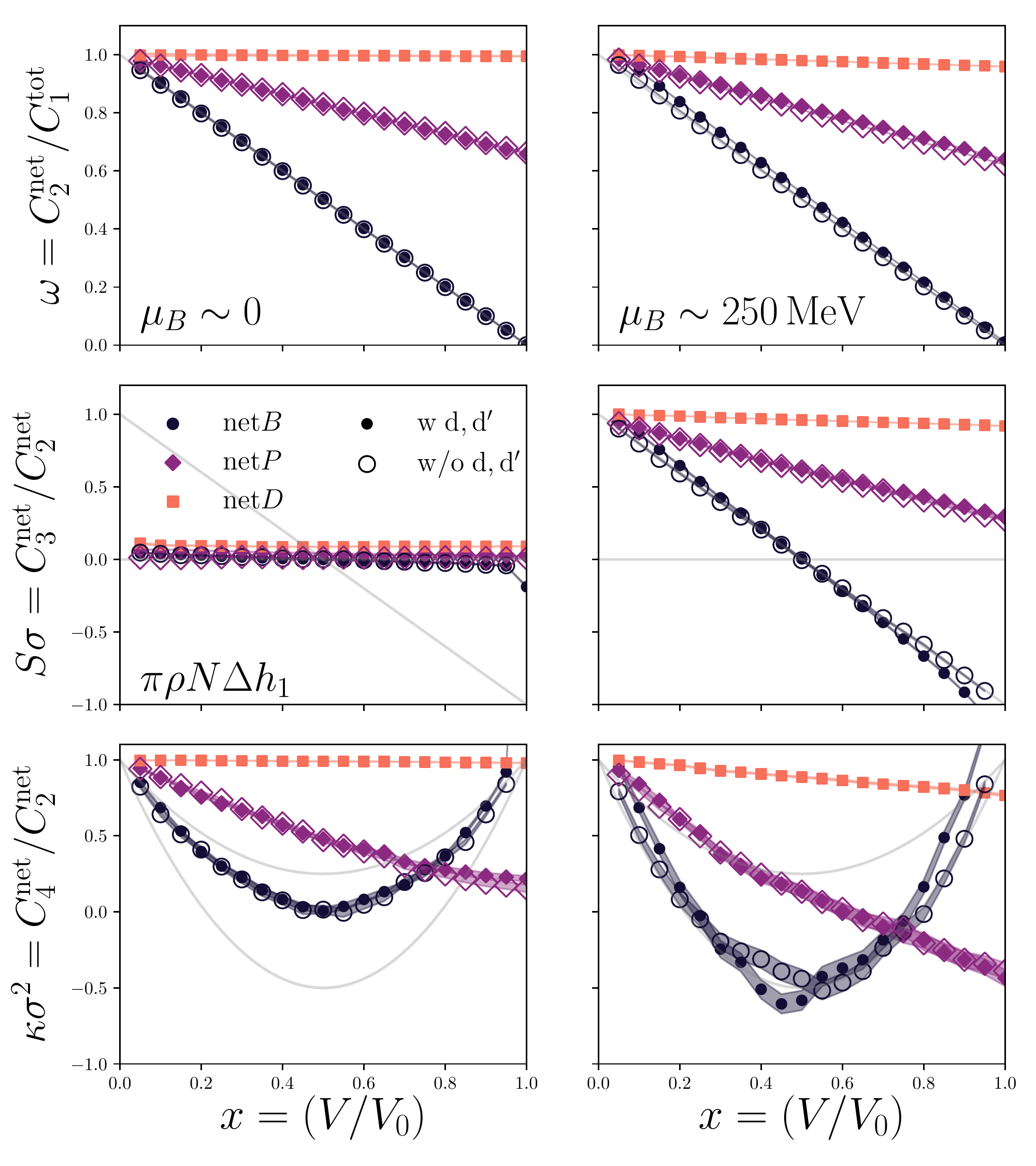} 
  \caption{Scaled variance (top), skewness (center) and kurtosis (bottom) as a function of the size of the subvolume. Open circles show the result of the box containing no deuterons whereas the closed circles correspond to results with deuteron formation.
           The results are presented for $\mu_B\sim 0$ (left column) and $\mu_B\sim 250\,\rm{MeV}$ (right column). Orange points show the deuteron number cumulants, purple points the proton cumulants and black points the baryon number cumulants.} 
  \label{CumulantsDeuterons}
\end{figure}

Fig. \ref{CumulantsDeuterons} shows the cumulants as a function of the subvolume size. For both systems 3 and 4, the baryon number cumulants follow the expected analytic conservation curves, which where observed before.
Interestingly in the case where deuterons are produced, the cumulants show no large dependence on the size of the subvolume, meaning they are rarely affected by baryon number conservation and are produced thermally. In addition the proton number cumulants of system 3 and 4 coincide with each other.
This means that the proton cumulants are not affected by the deuteron cluster formation and deuterons are rarely affected by conservation effects, since their yields are small and therefore they follow the thermal expectation.

\section{\label{conclusion}Summary and Conclusions}

In this work, we have studied the effects of conservation laws on fluctuation observables. Within a microscopic hadronic transport approach several hadronic systems have been evaluated in infinite matter calculations and different dynamic interactions have been investigated in detail. Resonance formation has no big effect on the higher moments for electric charge cumulants, as long as the density of particles is small enough. The kinematic cuts have always the expected effects on the cumulants of reducing the influence of charge conservation. For the net baryon number, that is of great interest in the context of identifying the critical endpoint in heavy-ion measurements, we have shown that baryon annihilation plays a bigger role at zero chemical potential and mainly affects the skewness and kurtosis. Interestingly, the proposed binomial mapping from net protons to net baryons suggested from isospin randomization cannot fully reconstruct the proper net baryon number cumulants. While for an artificial set of random protons the mapping works as expected, it does not for the actual SMASH protons containing correlations from the dynamic evolution. Last, the influence of cluster formation has been studied and the proton cumulants are largely unaffected. The deuterons themselves follow a thermal expectation unaffected by conservation laws. 
This is of relevance for the comparison of experimental results for fluctuation observables with theory calculations based on a grand canonical ensemble. In the future, it will be interesting to explore how the hadronic rescattering dynamics affects the cumulants in an expanding system. 

\begin{acknowledgments}
This work was supported by the DFG SinoGerman project (project number 410922684). Computational resources have been provided by the Center for Scientific Computing (CSC) at the Goethe-University of Frankfurt. H.E. acknowledges the support by the State of Hesse within the Research Cluster ELEMENTS (Project ID 500/10.006).
\end{acknowledgments}

\appendix
\section{Particles and interactions considered}\label{App:Systems}
Here a more detailed description of the particle content and interactions of the different calculations are presented. 
\begin{enumerate}
  \item[] System 1\\
      Particles: $\pi$, $\rho$ \\
      Interactions: $\rho\leftrightarrow\pi\pi$
  \item[] System 2\\
      Particles: $\pi$, $\rho$, $N$, $\Delta$ \\
      Interactions: $\rho\leftrightarrow\pi\pi$, $\Delta\leftrightarrow N\pi$
  \item[] System 3\\
      Particles: $\pi$, $\rho$, $N$, $\Delta$, $h_1(1170)$ \\
      Interactions: $\rho\leftrightarrow\pi\pi$, $\Delta\leftrightarrow N\pi$, $N\bar N\leftrightarrow 5\pi$
  \item[] System 4\\
      Particles: $\pi$, $\rho$, $N$, $\Delta$, $h_1(1170)$, $d$ $d^\prime$ \\
      Interactions: $\rho\leftrightarrow\pi\pi$, $\Delta\leftrightarrow N\pi$, $N\bar N\leftrightarrow 5\pi$, $Nnp\leftrightarrow Nd$
  \item[] System 5\\
      Full SMASH particles plus interactions

\end{enumerate}

\section{\label{toymodel}Monte Carlo Toy Model}
\begin{figure}[htp]
  \begin{center}
  \includegraphics[width=1.0\linewidth]{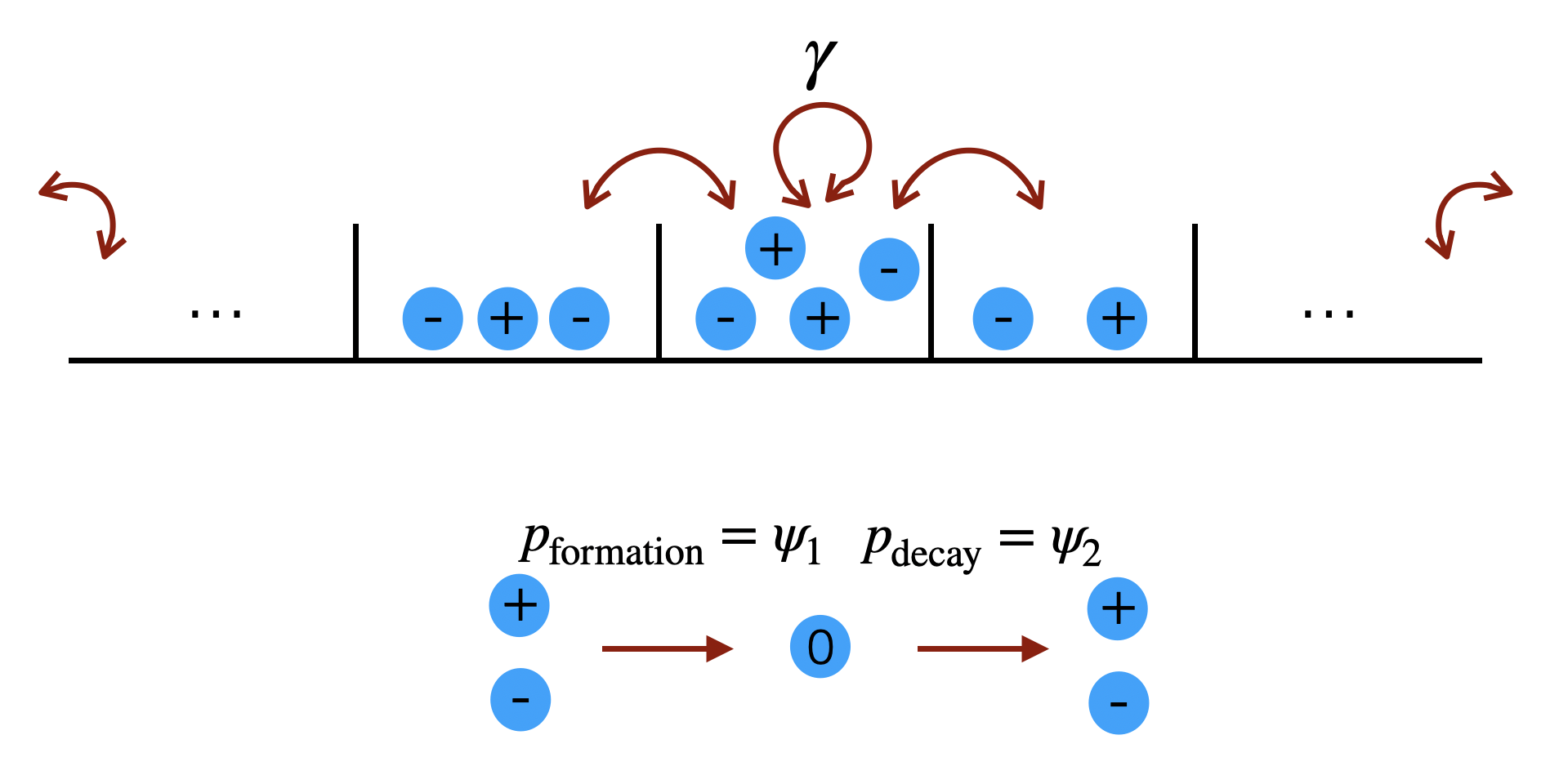} 
  \caption{Toy model to describe the influence of particle annihilation on the fluctuation of the net charge number.} 
  \label{ToyModel}
  \end{center}
\end{figure}
To better quantify the impact of charge annihilation on the net charge fluctuations, a simplistic Monte Carlo model is explored, see Fig. \ref{ToyModel}. On a one-dimensional grid, charged particles can move between each grid cell with a given probability $\gamma$ on a time-step basis. The probability is chosen to be $\gamma = 1/3$. This model so far is inspired by the diffusion master equation \cite{Kitazawa:2013bta}. In addition, the grid is build with periodic boundary conditions.

Besides the movement along the grid cells, a reaction is incorporated which annihilates a positive and negative charged particle and creates a particle with charge $0$. The probability of creating such a particle is given by the probability $\psi_1$. Per timestep, in each bin and for one set of combinations of positive and negative charged particles a particle of charge $0$ is created with the probability $\psi_1$ and the two opposite charged particles are removed from the bin.
If such a particle has been created, it can move between the bins and will eventually decay after some time with a probability $\rm{exp}(-t_{\rm{step}}\cdot \psi_2)$.
As a result, the net charge is conserved in the whole system and the number of total charge is controllable by the two parameters $\psi_1$ and $\psi_2$. 

After the system is initialized, it is evolved in time until it has reached chemical equilibrium. It is also checked that detailed balance in the system is fulfilled.
In the following, the equilibrium properties and the results of the cumulants of the described system are shown for three different sets of parameters.  The specific values are displayed in Tab. \ref{TablePsis}.

\begin{table}[hbt!]
  \begin{tabular}{l | c c}
               & $\psi_1$ & $\psi_2$ \\\hline
   Set 1 & 0.5 & 0.9 \\
   Set 2 & 0.1 & 0.95 \\
   Set 3 & 0.01 & 0.95
  \end{tabular}
  \caption{Sets of parameters $\psi_1$ and $\psi_2$ used in this work.}
  \label{TablePsis}
\end{table}

The parameters are chosen such that the final equilibrated value of $N_Q^{\rm{tot}}$ is different for all three sets.
In the initial state, only an equal number of $25$ positive and negative charged particles are randomly placed in a grid of $10$ bins.

\begin{figure}[htp]
  \centering
  \includegraphics[width=0.7\linewidth]{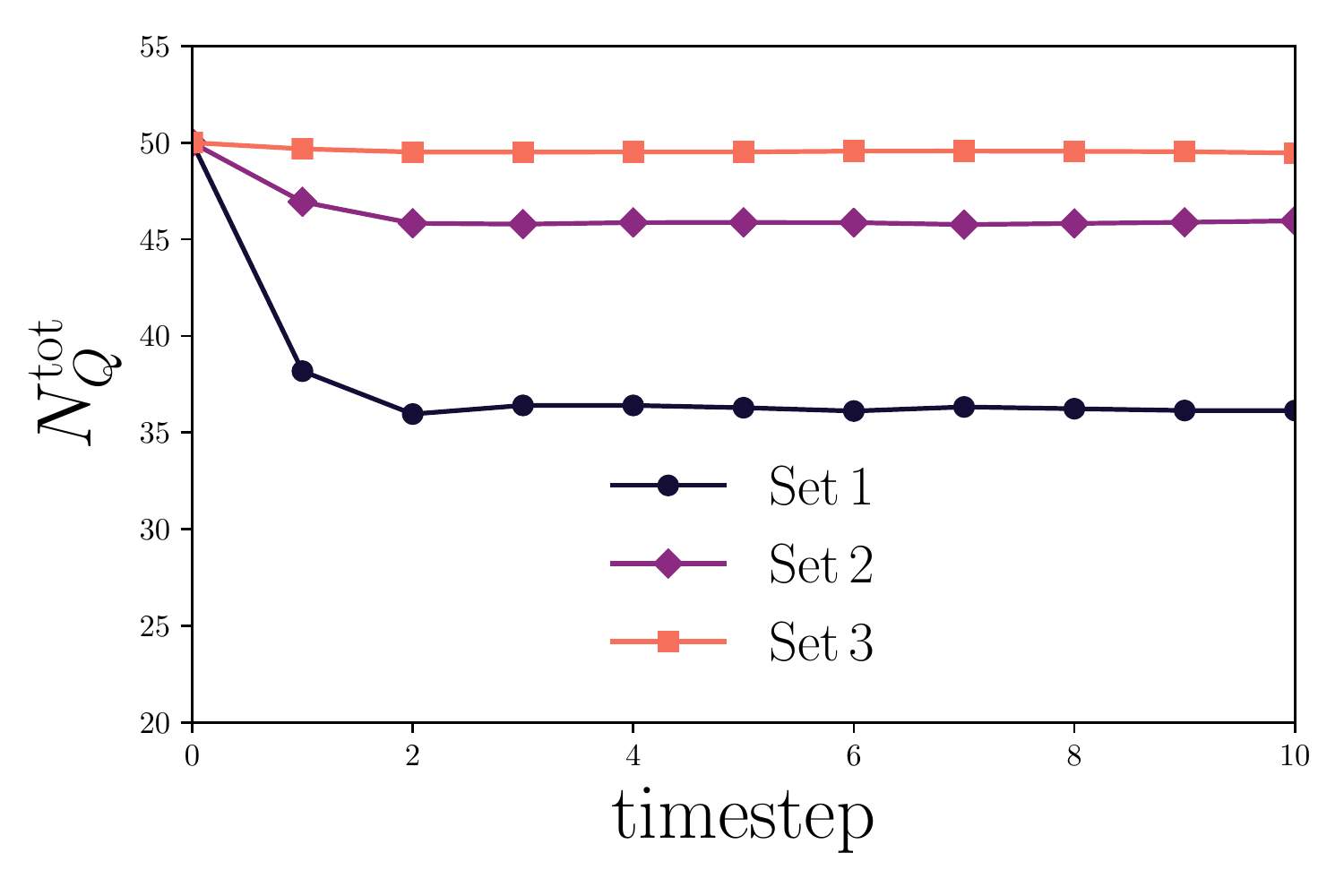} 
  \caption{Evolution of the total number of charge as a function of the timestep.} 
  \label{ToyModelQtotEvol}
\end{figure}
\begin{figure}[htp]
  \centering
  \includegraphics[width=0.7\linewidth]{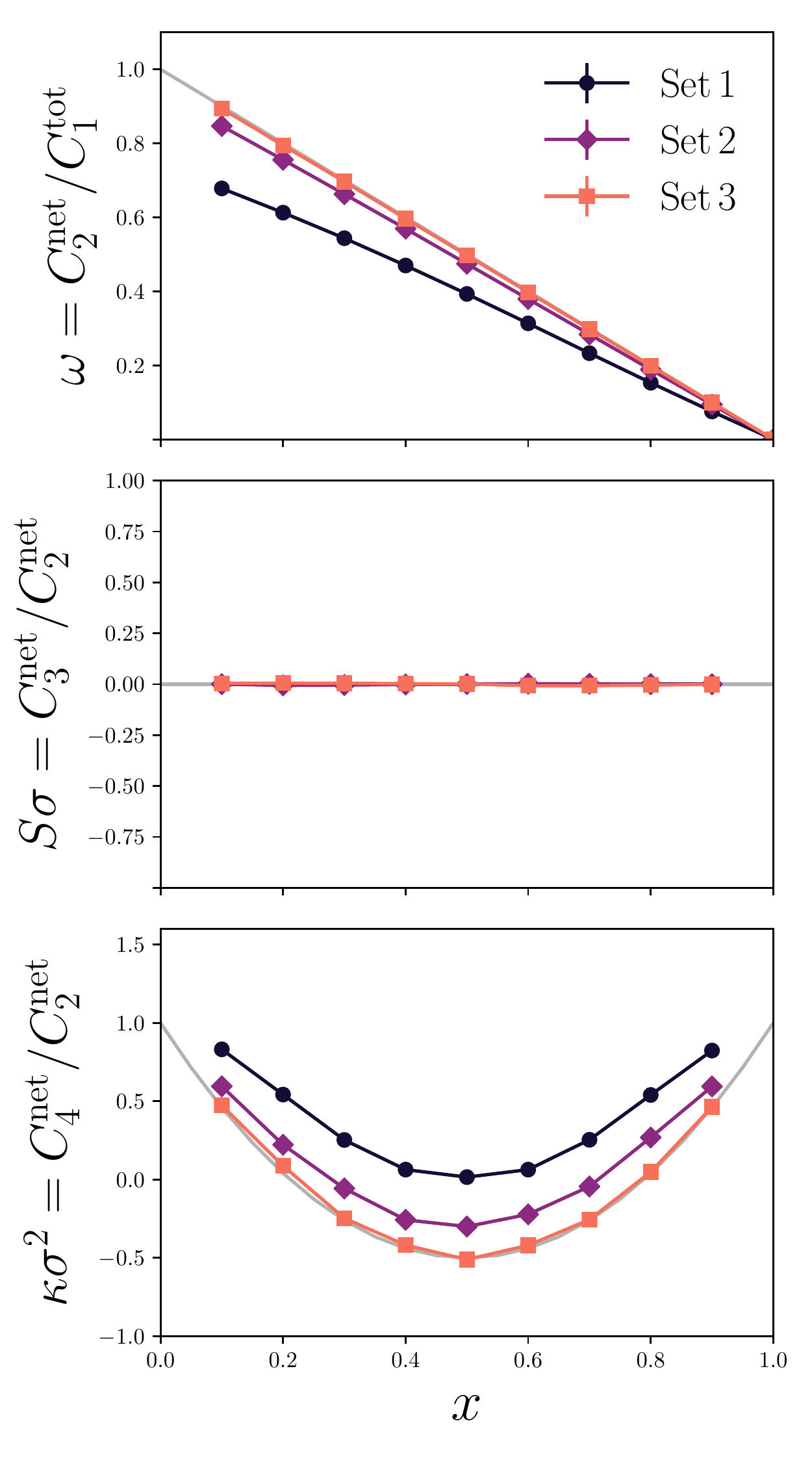} 
  \caption{Scaled variance (top) and kurtosis (bottom) as a function of $x$. The results for three different sets of parameter are shown for a system with a number of bins of $10$ and in total $50$ charged particles in the initial state.} 
  \label{ToyModelFluctuations}
\end{figure}

Fig. \ref{ToyModelQtotEvol} shows the evolution of the total number of charges in the system for the different sets of parameters $\psi_1$ and $\psi_2$. As can be seen, for the parameter set 1, there is a larger amount of neutral charged particles produced compared to the system with the parameter set 3.
As a result, $Q_{tot}$ equilibrates at a lower value compared to the parameter set 3, where only a small amount of particles with charge $0$ is produced, due to a reduced production probability.
To calculate event-by-event fluctuations in subvolumes by means of grid cells, many different events are calculated and the fluctuations are computed. Here, the content of each bin are used and summed up. E.g. in the case of 10 bins in total, the fluctuations of the net charge number is calculated from the sum of $9$ of those bins.  
Fig. \ref{ToyModelFluctuations} shows the scaled variance and the kurtosis for the three different sets of parameters as a function of $x$. Here, $x$ corresponds to the sum of bins used to calculate the fluctuations over the total number of bins.

Starting with the scaled variance, at lower values of $x$ the variance is reduced by a larger amount of charge $0$ particles in the system. For $x\rightarrow 1$, $\omega$ goes to zero, which is expected, since the net charge is conserved in the system.
For the parameter set 1, where only a very small amount of particles is annihilated, the fluctuations follow the expected conservation curve $1 - x$.
For the kurtosis, the same effect as presented in SMASH can be seen. For the parameter set 1, which creates a larger amount of particles with charge zero, the kurtosis is modified from the baseline of conservation $(1 - 6 x(1-x))$ and a strong effect can be seen around $x=0.5$.
This shows, that the effect observed in SMASH that an annihilation process modifies the kurtosis can be reproduced in a more simplistic model. This however is just a conceptual study. The results from SMASH are more realistic in terms of e.g. physical cross sections.

\bibliographystyle{ieeetr}
\bibliography{paper}

\begin{thebibliography}{10}

\bibitem{Asakawa:2015ybt}
M.~Asakawa and M.~Kitazawa, ``{Fluctuations of conserved charges in
  relativistic heavy ion collisions: An introduction},'' {\em Prog. Part. Nucl.
  Phys.}, vol.~90, pp.~299--342, 2016.

\bibitem{Stephanov:1999zu}
M.~A. Stephanov, K.~Rajagopal, and E.~V. Shuryak, ``{Event-by-event
  fluctuations in heavy ion collisions and the QCD critical point},'' {\em
  Phys. Rev. D}, vol.~60, p.~114028, 1999.

\bibitem{Hatta:2003wn}
Y.~Hatta and M.~A. Stephanov, ``{Proton number fluctuation as a signal of the
  QCD critical endpoint},'' {\em Phys. Rev. Lett.}, vol.~91, p.~102003, 2003.
\newblock [Erratum: Phys.Rev.Lett. 91, 129901 (2003)].

\bibitem{Bazavov:2017dus}
A.~Bazavov {\em et~al.}, ``{The QCD Equation of State to $\mathcal{O}(\mu_B^6)$
  from Lattice QCD},'' {\em Phys. Rev. D}, vol.~95, no.~5, p.~054504, 2017.

\bibitem{Borsanyi:2018grb}
S.~Borsanyi, Z.~Fodor, J.~N. Guenther, S.~K. Katz, K.~K. Szabo, A.~Pasztor,
  I.~Portillo, and C.~Ratti, ``{Higher order fluctuations and correlations of
  conserved charges from lattice QCD},'' {\em JHEP}, vol.~10, p.~205, 2018.

\bibitem{Stephanov:2006zvm}
M.~A. Stephanov, ``{QCD phase diagram: An Overview},'' {\em PoS}, vol.~LAT2006,
  p.~024, 2006.

\bibitem{Kumar:2013cqa}
L.~Kumar, ``{Review of Recent Results from the RHIC Beam Energy Scan},'' {\em
  Mod. Phys. Lett. A}, vol.~28, p.~1330033, 2013.

\bibitem{STAR:2020tga}
J.~Adam {\em et~al.}, ``{Nonmonotonic Energy Dependence of Net-Proton Number
  Fluctuations},'' {\em Phys. Rev. Lett.}, vol.~126, no.~9, p.~092301, 2021.

\bibitem{STAR:2021iop}
M.~Abdallah {\em et~al.}, ``{Cumulants and correlation functions of net-proton,
  proton, and antiproton multiplicity distributions in Au+Au collisions at
  energies available at the BNL Relativistic Heavy Ion Collider},'' {\em Phys.
  Rev. C}, vol.~104, no.~2, p.~024902, 2021.

\bibitem{Vovchenko:2021gas}
V.~Vovchenko, ``{Phenomenological developments for event-by-event fluctuations
  of conserved charges},'' in {\em {International Conference on Critical Point
  and Onset of Deconfinement}}, 10 2021.

\bibitem{Koch:2008ia}
V.~Koch, {\em {Hadronic Fluctuations and Correlations}}, pp.~626--652.
\newblock 2010.

\bibitem{Bzdak:2017ltv}
A.~Bzdak and V.~Koch, ``{Rapidity dependence of proton cumulants and
  correlation functions},'' {\em Phys. Rev. C}, vol.~96, no.~5, p.~054905,
  2017.

\bibitem{Bzdak:2012an}
A.~Bzdak, V.~Koch, and V.~Skokov, ``{Baryon number conservation and the
  cumulants of the net proton distribution},'' {\em Phys. Rev. C}, vol.~87,
  no.~1, p.~014901, 2013.

\bibitem{Vovchenko:2020tsr}
V.~Vovchenko, O.~Savchuk, R.~V. Poberezhnyuk, M.~I. Gorenstein, and V.~Koch,
  ``{Connecting fluctuation measurements in heavy-ion collisions with the
  grand-canonical susceptibilities},'' {\em Phys. Lett. B}, vol.~811,
  p.~135868, 2020.

\bibitem{Nahrgang:2009dqc}
M.~Nahrgang, T.~Schuster, M.~Mitrovski, R.~Stock, and M.~Bleicher,
  ``{Net-baryon-, net-proton-, and net-charge kurtosis in heavy-ion collisions
  within a relativistic transport approach},'' {\em Eur. Phys. J. C}, vol.~72,
  p.~2143, 2012.

\bibitem{Kuznietsov:2022pcn}
V.~A. Kuznietsov, O.~Savchuk, M.~I. Gorenstein, V.~Koch, and V.~Vovchenko,
  ``{Critical point particle number fluctuations from molecular dynamics},'' 1
  2022.

\bibitem{Petersen:2015pcy}
H.~Petersen, D.~Oliinychenko, J.~Steinheimer, and M.~Bleicher, ``{Influence of
  kinematic cuts on the net charge distribution},'' {\em Nucl. Phys. A},
  vol.~956, pp.~336--339, 2016.

\bibitem{Kitazawa:2012at}
M.~Kitazawa and M.~Asakawa, ``{Relation between baryon number fluctuations and
  experimentally observed proton number fluctuations in relativistic heavy ion
  collisions},'' {\em Phys. Rev. C}, vol.~86, p.~024904, 2012.
\newblock [Erratum: Phys.Rev.C 86, 069902 (2012)].

\bibitem{Feckova:2015qza}
Z.~Feckov\'a, J.~Steinheimer, B.~Tom\'a\v{s}ik, and M.~Bleicher, ``{Net-proton
  number kurtosis and skewness in nuclear collisions: Influence of deuteron
  formation},'' {\em Phys. Rev. C}, vol.~92, no.~6, p.~064908, 2015.

\bibitem{Weil:2016zrk}
J.~Weil {\em et~al.}, ``{Particle production and equilibrium properties within
  a new hadron transport approach for heavy-ion collisions},'' {\em Phys. Rev.
  C}, vol.~94, no.~5, p.~054905, 2016.

\bibitem{SMASH_github}
\url{https://smash-transport.github.io}.
\newblock Accessed: 2022-02-10.

\bibitem{dmytro_oliinychenko_2020_4336358}
D.~Oliinychenko, V.~Steinberg, J.~Weil, J.~Staudenmaier, M.~Kretz, A.~Schäfer,
  H.~E. (Petersen), S.~Ryu, J.~Rothermel, J.~Mohs, F.~Li, A.~Sorensen,
  D.~Mitrovic, L.~Pang, J.~Hammelmann, A.~Goldschmidt, M.~Mayer,
  O.~Garcia-Montero, N.~Kübler, and Nikita, ``smash-transport/smash:
  Smash-2.0,'' Dec. 2020.

\bibitem{Staudenmaier:2020xqr}
J.~Staudenmaier, N.~K\"ubler, and H.~Elfner, ``{Particle production in AgAg
  collisions at $E_{\rm Kin}=1.58A$ GeV within a hadronic transport
  approach},'' {\em Phys. Rev. C}, vol.~103, no.~4, p.~044904, 2021.

\bibitem{Mohs:2019iee}
J.~Mohs, S.~Ryu, and H.~Elfner, ``{Particle Production via Strings and Baryon
  Stopping within a Hadronic Transport Approach},'' {\em J. Phys. G}, vol.~47,
  no.~6, p.~065101, 2020.

\bibitem{Steinberg:2019wgm}
V.~Steinberg, J.~Steinheimer, H.~Elfner, and M.~Bleicher, ``{Constraining
  resonance properties through kaon production in pion\textendash{}nucleus
  collisions at low energies},'' {\em J. Phys. G}, vol.~48, no.~2, p.~025109,
  2021.

\bibitem{Rose:2017bjz}
J.~B. Rose, J.~M. Torres-Rincon, A.~Sch\"afer, D.~R. Oliinychenko, and
  H.~Petersen, ``{Shear viscosity of a hadron gas and influence of resonance
  lifetimes on relaxation time},'' {\em Phys. Rev. C}, vol.~97, no.~5,
  p.~055204, 2018.

\bibitem{Hammelmann:2018ath}
J.~Hammelmann, J.~M. Torres-Rincon, J.-B. Rose, M.~Greif, and H.~Elfner,
  ``{Electrical conductivity and relaxation via colored noise in a hadronic
  gas},'' {\em Phys. Rev. D}, vol.~99, no.~7, p.~076015, 2019.

\bibitem{Rose:2020sjv}
J.-B. Rose, M.~Greif, J.~Hammelmann, J.~A. Fotakis, G.~S. Denicol, H.~Elfner,
  and C.~Greiner, ``{Cross-conductivity: novel transport coefficients to
  constrain the hadronic degrees of freedom of nuclear matter},'' {\em Phys.
  Rev. D}, vol.~101, no.~11, p.~114028, 2020.

\bibitem{Manley:1992yb}
D.~M. Manley and E.~M. Saleski, ``{Multichannel resonance parametrization of pi
  N scattering amplitudes},'' {\em Phys. Rev. D}, vol.~45, pp.~4002--4033,
  1992.

\bibitem{ParticleDataGroup:2020ssz}
P.~A. Zyla {\em et~al.}, ``{Review of Particle Physics},'' {\em PTEP},
  vol.~2020, no.~8, p.~083C01, 2020.

\bibitem{Bluhm:2016byc}
M.~Bluhm, M.~Nahrgang, S.~A. Bass, and T.~Schaefer, ``{Impact of resonance
  decays on critical point signals in net-proton fluctuations},'' {\em Eur.
  Phys. J. C}, vol.~77, no.~4, p.~210, 2017.

\bibitem{Luo:2017faz}
X.~Luo and N.~Xu, ``{Search for the QCD Critical Point with Fluctuations of
  Conserved Quantities in Relativistic Heavy-Ion Collisions at RHIC : An
  Overview},'' {\em Nucl. Sci. Tech.}, vol.~28, no.~8, p.~112, 2017.

\bibitem{Luo:2015ewa}
X.~Luo, ``{Energy Dependence of Moments of Net-Proton and Net-Charge
  Multiplicity Distributions at STAR},'' {\em PoS}, vol.~CPOD2014, p.~019,
  2015.

\bibitem{Karsch:2015zna}
F.~Karsch, K.~Morita, and K.~Redlich, ``{Effects of kinematic cuts on
  net-electric charge fluctuations},'' {\em Phys. Rev. C}, vol.~93, no.~3,
  p.~034907, 2016.

\bibitem{Garcia-Montero:2021haa}
O.~Garcia-Montero, J.~Staudenmaier, A.~Sch\"afer, J.~M. Torres-Rincon, and
  H.~Elfner, ``{The role of proton-antiproton regeneration in the late stages
  of heavy-ion collisions},'' 7 2021.

\bibitem{Savchuk:2021aog}
O.~Savchuk, V.~Vovchenko, V.~Koch, J.~Steinheimer, and H.~Stoecker,
  ``{Constraining baryon annihilation in the hadronic phase of heavy-ion
  collisions via event-by-event fluctuations},'' 6 2021.

\bibitem{Staudenmaier:2021lrg}
J.~Staudenmaier, D.~Oliinychenko, J.~M. Torres-Rincon, and H.~Elfner,
  ``{Deuteron production in relativistic heavy ion collisions via stochastic
  multiparticle reactions},'' {\em Phys. Rev. C}, vol.~104, no.~3, p.~034908,
  2021.

\bibitem{Demir:2008tr}
N.~Demir and S.~A. Bass, ``{Shear-Viscosity to Entropy-Density Ratio of a
  Relativistic Hadron Gas},'' {\em Phys. Rev. Lett.}, vol.~102, p.~172302,
  2009.

\bibitem{Feckova:2016kjx}
Z.~Feckov\'a, J.~Steinheimer, B.~Tom\'a\v{s}ik, and M.~Bleicher, ``{Formation
  of deuterons by coalescence: Consequences for deuteron number
  fluctuations},'' {\em Phys. Rev. C}, vol.~93, no.~5, p.~054906, 2016.

\bibitem{Oliinychenko:2018ugs}
D.~Oliinychenko, L.-G. Pang, H.~Elfner, and V.~Koch, ``{Microscopic study of
  deuteron production in PbPb collisions at $\sqrt{s} = 2.76 TeV$ via
  hydrodynamics and a hadronic afterburner},'' {\em Phys. Rev. C}, vol.~99,
  no.~4, p.~044907, 2019.

\bibitem{Oliinychenko:2020znl}
D.~Oliinychenko, C.~Shen, and V.~Koch, ``{Deuteron production in AuAu
  collisions at $\sqrt{s_{NN}}=$7\textendash{}200 GeV via pion catalysis},''
  {\em Phys. Rev. C}, vol.~103, no.~3, p.~034913, 2021.

\bibitem{Kitazawa:2013bta}
M.~Kitazawa, M.~Asakawa, and H.~Ono, ``{Non-equilibrium time evolution of
  higher order cumulants of conserved charges and event-by-event analysis},''
  {\em Phys. Lett. B}, vol.~728, pp.~386--392, 2014.

\end{thebibliography}

\end{document}